\documentstyle[11pt,aaspp4,flushrt]{article}
\newcommand\etal{{\it et al.\/}}

\newcommand\hn{{H$_0$}}
\newcommand\kthirty{{\bar\kappa_{30^{\prime\prime}}}}
\newcommand\hunits{${\rm km}\, {\rm s}^{-1}\, {\rm Mpc}^{-1}$}
 
\begin{document}
\renewcommand{\topfraction}{0.8}
\renewcommand{\textfraction}{0.15}
 
\slugcomment{Accepted by {\it The Astronomical Journal} 3/16/99}
 
\title{Values of \hn\ from Models of the Gravitational Lens $0957+561$}
 
\author{Gary Bernstein and Philippe Fischer\footnote{Hubble Fellow}}
\affil{Dept. of Astronomy, University of Michigan, Ann Arbor, MI
48109; garyb,~philf@astro.lsa.umich.edu}
 
\begin{abstract}
The lensed double QSO
0957+561 has a well-measured time delay and hence is useful for
a global determination of \hn.  Uncertainty in the mass distribution
of the lens is the largest source of uncertainty in the derived \hn.
We investigate the range of \hn\ produced by a set of lens models
intended to mimic the full range of astrophysically plausible mass
distributions, using as constraints the numerous multiply-imaged 
sources which have been detected.  We obtain the first adequate fit to 
all the observations, but only if we include effects from the galaxy
cluster beyond
a constant local magnification and shear.  Both the lens galaxy and
the surrounding cluster must depart from circular symmetry as well.

Lens models which are consistent with observations to 95\% CL indicate
$H_0=104^{+31}_{-23}(1-\kthirty)$ \hunits.
Previous weak lensing
measurements constrain the mean mass density within 30\arcsec\ of G1
to be $\kthirty=0.26\pm0.16$ (95\% CL), implying 
$H_0=77^{+29}_{-24}$ \hunits\ (95\% CL). The best-fitting models
span the range 65--80 \hunits.
Further observations will
shrink the confidence interval for both the mass model and
$\kthirty$. 

The range of \hn\ allowed by the full gamut of our lens models is
substantially 
larger than that implied by limiting consideration to simple power
law density profiles.  We therefore caution against use of
simple isothermal or power-law mass models in the derivation of \hn\
from other time-delay systems.  High-S/N imaging of multiple or
extended lensed features will greatly reduce the \hn\ uncertainties when
fitting complex models to time-delay lenses.

\end{abstract}
 
\keywords{distance scale---gravitational lensing---galaxies:
elliptical---dark matter}
\section{Motivations}
 
The accurate measure of the time delay between the two images of the
gravitationally lensed quasar Q0957+561
(Kundi\'c \etal\markcite{Ku1} 1996) leads, in principle, to a
measure of \hn\ accurate to a few percent (Refsdal
\markcite{Re1} 1964).  This accuracy in \hn\ is achievable only when
the gravitational potential $\phi$ of the lens (or, equivalently, its 
surface mass distribution $\Sigma$) 
is also determined to an accuracy of a few percent.  At
present, uncertainties in the lens model dominate the uncertainty in
\hn.  The goal of this paper is to determine which, if any, models for
the lens are consistent with the many observational constraints on
this system, and to thence find the range of \hn\ values implied by
the family of acceptable lens models.

A number of authors have investigated mass models for $0957+561$ over the
20 years since its discovery.  
Gorenstein, Falco, \& Shapiro\markcite{Go3} (1988a; followed by
Falco, Gorenstein, \& Shapiro\markcite{Fa1} 1991, hereafter
FGS) pointed out that any lens model constrained by the positions or
magnifications of lensed objects are subject to a ``mass sheet
degeneracy.'' If a mass distribution $\Sigma(\vec x)$ successfully
reproduces the lensing behavior, then an altered model with mass
distribution $(1-\kappa)\Sigma(\vec x) + \kappa\Sigma_{\rm crit}$ will
have identical optical characteristics yet yield a value for \hn\
differing by a factor $1-\kappa$.  Since the primary lensing
galaxy G1 is the brightest member of a modest galaxy cluster, we fully
expect there to be a dark matter component which is slowly varying
across the region of multiple imaging.  The lens modeling process can
hence be broken into two fairly distinct problems: first, we must
determine a ``strong lensing model'' which prescribes a mass
distribution $\Sigma_0$, over the central $\approx10\arcsec$ region,
that accurately reproduces the observed strongly distorted or multiply
imaged sources in this area.  Second, we must find a way to measure the
average mass density across the strong-lensing area or otherwise find
the proper $1-\kappa$ scaling for $\Sigma_0$ (and hence \hn).

The primary focus of this paper is to investigate several broad
classes of candidate mass distributions for the G1-cluster system, and
find what range of \hn\ is produced by those which satisfy the
observational constraints on the lens.  In \S\ref{constraints} we
delineate these constraints, which are now more extensive than
available to Grogin \&
Narayan \markcite{Gr1} (1996, hereafter GN).  In
\S\ref{models} we describe the parametric models of the mass
distribution which we will use to fit the strong-lensing constraints,
and the {\it a priori} constraints we impose on these models in order
to keep them ``astrophysically reasonable.''  In \S\ref{fits} we
present the results of the fits, including the best-fit models and the
range of \hn\ allowed {\it before} application of the $1-\kappa$
correction.  \S\ref{constraints}--\ref{fits} are somewhat involved,
and many readers may wish to skip through to the end of \S\ref{fits}
for a summary of the model-fitting results.
In \S\ref{weak} we use the weak-lensing
measurements of Fischer \etal\markcite{Fi1} (1996, hereafter Paper I)
to resolve the mass-sheet degeneracy.  In \S\ref{improve} we discuss
the possibilities for improvement in the
constraints on \hn\ that we derive.
In \S\ref{conclusion} we
conclude.

\section{Strong-Lensing Constraints}
\label{constraints}
The $0957+561$ system has been observed in detail at wavelengths from
radio to x-ray, and displays a rich variety of lensed features.  We
discuss here the numerical constraints that these observations place
upon models of the lens optics.  Table~\ref{tconstr} 
summarizes the constraints we
have adopted for our modeling.

\begin{deluxetable}{cllc}
\tablewidth{0pt}
\tablecaption{Adopted Constraints on Lens Models}
\tablehead{
\colhead{Object} &
\colhead{$x$\tablenotemark{1}} & \colhead{$y$\tablenotemark{1}} &
\colhead{Flux Ratio}
}
\startdata
Quasar A & $+1.408(0)$\tablenotemark{2} & $+5.034(0)$\tablenotemark{2} 
	& $f_{QB}/f_{QA}=0.74(2)$ \\
Quasar B & $+0.182(0)$\tablenotemark{2} & $-1.018(0)$\tablenotemark{2}
	& \\[6pt]
Jet A5 & $+0.0164(5)$\tablenotemark{3} 
	& $+0.0457(5)$\tablenotemark{3} 
	& $f_{B5}/f_{A5}=0.63(3)$ \\
Jet B5 & $+0.0181(5)$\tablenotemark{3} 
	& $+0.0559(5)$\tablenotemark{3} 
	& \\[6pt]
Knot 1 &  $+0.06(5)$ & $-2.55(5)$ 
	& $\ln(f_{K2}/f_{K1})=0.0(7)$ \\
Knot 2 &  $+0.48(5)$ & $-2.43(5)$ 
	& \\[6pt]
Blob 2 & $-1.54(5)$ & $-0.05(5)$ 
	& $\ln(f_{B3}/f_{B2})=0.9(3)$ \\
Blob 3 & $+2.86(5)$ & $+3.47(5)$ & 
\enddata
\tablecomments{Image positions and flux ratios for all 4 pairs of
multiple images are given, with 1-sigma uncertainties in parentheses.
All sources assumed at $z=1.41$ and lens assumed at $z=0.356$.  See
\S2 for details and references.}
\tablenotetext{1}{Positions given in arcseconds relative to G1 center,
with $x$ pointing West and $y$ to North in J2000}
\tablenotetext{2}{Quasar positions are taken as exact.}
\tablenotetext{3}{Jet positions are relative to quasar cores.}
\label{tconstr}
\end{deluxetable}

\subsection{Quasar and G1 Positions: 2 constraints}
The positions of the A and B quasar cores are determined to
micro-arcsecond accuracy by the VLBI measurements of Gorenstein
\etal\markcite{Go2} (1988b).  We will adopt their positions and consider
them to be known exactly.  The requirement of a common source position
for the A and B quasar images places two (exact) constraints on the lens
model.  In our modeling procedure, these are solved by adjusting the
two components of external shear (see \S\ref{methods}).

The WFPC2 observations of Bernstein \etal\markcite{Be1}
(1997, hereafter Paper II) showed that the optical quasar separation
agrees with the VLBI separation to the measurement precision of a few
milli-arcseconds.  Paper II also shows that the optical peak and
centroid of G1 
coincide with the VLBI point source G$^\prime$ (Gorenstein \etal\ 1988b) 
to within 10~mas.  We will therefore adopt the VLBI position of
G$^\prime$ as the center of G1, and consider its 1~mas uncertainty to
be negligible.  We adopt the
G$^\prime$ position as the center of our mass distributions, so the G1
center does not appear as an adjustable parameter in our models.

We will henceforth measure all object positions in a coordinate system
centered on G$^\prime$, with $x$ axis pointing West and $y$ axis North
(J2000), with units of arcseconds unless otherwise specified.
Position angles will be measured counter-clockwise from the $x$-axis,
90\arcdeg\ different from the astronomical North-through-East convention.

\subsection{Quasar Jets: 2 Constraints}
The most detailed images to date of the VLBI jets extending from the A
and B quasars are given by Garrett \etal\markcite{Ga2} (1994), and are
re-analyzed by Barkana \etal\markcite{Ba1} (1998, hereafter BLFGKS).
Both authors fit to each jet a model containing a core and five 
additional Gaussian jet
components (A2--A6, B2--B6).  A proper lens model
should map each of these 5 pairs of objects to common sources.  Each
of these papers derives a local transformation which maps
the A jet positions and fluxes into their B counterparts.  To
simultaneously fit the positional constraints and the flux
magnification constraints (see the next section), it is necessary to
allow this map to be more complex than a linear
transformation.  Both papers fit a model in which the relative
magnification matrix is allowed to vary in a limited fashion along the
jet ({\it e.g.\/} the magnification eigenvectors are fixed but the
eigenvalues vary along the jet).  We have, therefore, two choices in
implementing the jet constraints in our models.  We can either fit to
some or all of the jet positions and fluxes directly, or we can fit by
trying to match the 6 derived parameters that describe the local
behavior of the relative magnification matrix.  We choose the former
method for two reasons.  First, computing the gradient of the local magnification
matrix would require computing complicated ratios of third derivatives
of the model potential, which would slow our numerical methods
substantially.  Second, to reduce the behavior of the local
magnification matrix to 6 parameters, Garret \etal\ and BLFGKS assume
that the eigenvectors of the magnification do not vary with position.
This may not be the case for our model lenses, and it is not clear
how, in this case,  we should
implement a fit to the parameters of a fixed-eigenvector model.

Fitting directly to the jet component positions 
could add 10 constraints to our model, but in fact most of this
information is redundant for any realistic model (flux ratio
constraints are discussed below).
The jet
components lie nearly along a line, and their positions
are consistent with a constant
relative magnification matrix between the A and B images. The
nearly-linear arrangement of the sources further means that only two
components of the relative magnification matrix are well constrained.
We can therefore extract nearly all the useful information from the
jet components by considering only the brightest and best-measured
pair, A5 and B5.  We will use the positions from the ``partial fit''
of BLFGKS (their Table~2), for which in fact all the jet positions are
forced to map smoothly from A to B.

The positions of components A5 and B5 are determined to high
precision, with formal uncertainties of only $\approx0.1$~mas, or
0.2\% of their displacements from the quasar cores.  For the reasons
outlined in Appendix~\ref{jets}, we believe that the use of such small
uncertainties may be unjustified and/or could constrain the $\chi^2$
minimum to a misleadingly narrow region of parameter space. When
fitting models, we give the models more freedom by widening 
the error ellipse for each jet
component to be a circle with radius equal to 1\% of the jet's
displacement from the quasar cores.  In actuality we find that the
models do not require this additional freedom:
retaining the original BLFGKS error estimates changes the minimum $\chi^2$ by
$<0.1$ and changes the \hn\ bounds by $<1\%$.

\subsection{Quasar/Jet Flux Ratios:  2 Constraints}
Determination of the flux ratios between the A and B images is
confounded by three effects:  first, small sources can vary on time scales
comparable to the 1.1-year time delay.  Second, the quasar
continuum source is likely small enough to be microlensed by stars in
G1, which can cause decades-long perturbations to the flux ratio.
Third, Connor \etal\markcite{Co2} (1992) argue that the flux ratio 
varies significantly along the jet, as the core is closer to the
lensing caustic than the jets.

Garrett \etal\ (1994) summarize the various constraints on flux
ratios.  The components of the jet should be large enough to be free
of microlensing and temporal variation problems.  Their measured
flux ratio for B5/A5 combined with the previous independent VLBI jet flux
ratio measurements give a flux
ratio of $0.63\pm0.03$ at the position of jet component 5.  We also
know (from the jet images) that there must be a parity flip between
images A and B.

Measuring the flux ratio at the core is more difficult because
microlensing and time variation are likely.  
Garrett \etal\ (1994) cite several attempts to measure
the core flux ratio in the radio by interpolating observations at
different epochs and/or frequencies to make up for the 1.1-year time
delay.  All of these, however, are imprecise or used a time delay
value now excluded by the data.  Schild \& Smith (1991\markcite{Sc2})
measured broad-line \ion{Mg}{2}
fluxes from the two quasars at two epochs 1.1~years apart.  The
line flux originates from a region believed large enough to be
unaffected by microlensing, and the two epochs serve to remove any
source variability.  They report a flux ratio of $0.75\pm0.02$.
Spectrophotometric observations such as this are subject to many
systematic difficulties, and the quoted errors encompass only
counting statistics, so we will be
conservative and double the quoted uncertainty on this value.

More recently Haarsma \etal\markcite{Ha1} (1999) estimate a core flux
ratio from a long VLA time series.  The VLA does not resolve the core
from the jets, but if one presumes that the jets are invariant on
decade time scales, then the ratio of {\it fluctuations} in the A
and B fluxes (when phased by the time delay), 
gives a magnification ratio for the (varying) core flux
only.  The core radio continuum source is believed to be large enough
to avoid microlensing amplification. Analyses of different frequencies
and subsets of the data yield 
flux ratios from 0.72 to 0.76; we will therefore adopt $0.74\pm0.02$
as the core flux ratio, which is consistent with all optical and VLA
measurements.

Most recently BLFGKS derive core and Jet~5 flux ratios of
$0.74\pm0.06$ and $0.64\pm0.03$, respectively, from their ``partial
fit'' to the positions and fluxes of the VLBI jets.  This is
comfortingly consistent with the values we have adopted from other
sources (though the data upon which the jet flux ratio is based is the
same as the Garrett \etal\ data).

\subsection{Arc System: 6 Constraints}
Paper II gives the positions and fluxes of a number of faint objects
discovered in the strong-lensing region in WFPC2 images.  Three
resolved objects, ``Blob 2,'' ``Blob3,'' and an apparent arc, are
close enough to G1 to expect that they are multiply imaged.  The arc
contains two bright spots, a pattern which suggest that these
``Knots'' sit astride the critical line and are multiple images of a
bright spot in the source.  We adopt the positions and uncertainties
of Knots 1 and 2 given in Paper II.  In addition, we demand that Knots 1
and 2 have opposite parities, and that their flux ratio be
$\ln(f_{K1}/f_{K2})=0\pm0.7$, in accord with the (poorly) determined
magnitudes from Paper II.  This adds 3 constraints to the model.
We also enforce the qualitative constraint that a model must ``fold''
the source of the arc back over on itself, {\it i.e.\/} we expect that
the arc is a greatly magnified image of a small source, rather than an
image of some intrinsically very elongated source.  In practice this
qualitative constraint forces the G1 matter distribution to have a
position angle (PA) roughly aligned with the visible galaxy rather
than perpendicular to it (see \S\ref{fits}).  This conclusion is in
line with that of Keeton, Kochanek, \& Falco (1998\markcite{Ke1}), who
find that the application of isothermal ellipsoid mass models to 17
well-measured lens systems yeilds a position angle within
$\approx10\arcdeg$ of the visible $PA$ in nearly all cases.

Any reasonable model for the lens shows that the sources of Blobs 2
and 3 could be quite close to each other.  It is also clear that if
either source is at a redshift $z\gtrsim0.5$, then a counterimage
should be visible.  The only candidate counterimage for Blob 2 is Blob
3, and vice-versa, so there seems to be little risk in assuming these
two objects to be images of the same source.  We adopt this constraint
in our models, using the positions and uncertainties from Paper II.
We also constrain the flux ratio of Blob 2 to Blob 3 to be
$\ln(f_{B2}/f_{B3})=-0.9\pm0.3$.  This implies a $1.0\pm0.3$~mag
difference between the Blobs, which differs a bit from the
$1.3\pm0.14$~mag difference in Table~2 of Paper II.  We have revised
our estimate of the magnitude of Blob~3 by 0.3~mag.  The revision
reflects a changed estimate of the local sky value, which is difficult
to evaluate due to residual flux from the quasar and G1 images.

The magnification ratio will change rapidly across the extent of the
blobs, so their observed flux ratio is actually an integral of the
magnification ratio across the extent of the source.  Such a
calculation is impractical for our models (the shape of the source is
not well known anyway), so our constraint is only upon the
magnification ratio at the object centroids.  For this reason the
uncertainty on the magnification ratio used to constrain the model is
higher (0.3~mag) than the stated measurement uncertainties on the
relative flux (0.14~mag).  With improved S/N on the images of the
Blobs it should be possible to produce more specific constraints (and
better object names).

The two multiply-imaged systems provide 3 constraints each (2
position, 1 flux) to the model.  There are, however, two additional
degrees of freedom which are introduced, namely the redshifts of the
arc and blob source objects.  If we leave these source redshifts as
free parameters in the model fits described below, we find the best
fits have the arc and blob sources both very close to the quasar in
redshift.  Avruch \etal\markcite{Av1} (1997) obtain the same result,
though BLFGKS differ.  Furthermore, the arc and blob
sources are separated by only a few tenths of an arcsecond in the
source plane in these models.  
This strongly suggests that (a) the arc is a highly
magnified double image of an extension of the blob source that crosses
a caustic, and hence the arc and blob have common redshift; (b)
the arc/blob source object is at the quasar redshift.  The {\it a
priori} most likely distance for the arc and blob sources is of course
near the quasar, since quasars are found in large galaxies and since
galaxies are clustered in space.

More evidence that the arc and blob sources are at the quasar redshift
is given by a recent HST/NICMOS $H$-band image of this system
(Kochanek \etal\markcite{Ko2} 1998), which shows a spectacular pair of
arcs surrounding each quasar image, presumably the images of the
quasar host galaxy.  The WFPC2 $V$-band Blobs 2 and 3 and arc all lie
within the envelope of the NICMOS arcs.  It thus seems extremely
likely that the Blob and WFPC2 arc sources are ``hot spots,'' bright
in rest-frame UV, within or associated with the quasar host galaxy.
We will therefore assume a common redshift for all these sources.

\subsection{Other Information}
\subsubsection{Arcs}
Among the many interesting features revealed in the wealth of imaging
of the $0957+561$ system are extended arc-like structures seen in
NICMOS near-IR images (Kochanek \etal\ 1998), in high-S/N radio
imaging (Harvanek \etal\markcite{Ha2} 1997; Porcas
\etal\markcite{Po1} 1996, Avruch \etal\ 1997), and perhaps even in
x-ray images (Jones \etal\markcite{Jo1} 1993, Chartas
\etal\markcite{Ch2} 1995); It is difficult to
incorporate this information into our lens models unless the images
have sufficiently high S/N and resolution that one can identify a
correspondence between multiply imaged features in the
surface-brightness maps.  The x-ray ``arcs'' are a marginal detection
and not yet useful as lens model constraints.

Avruch \etal\ (1997) demonstrate that the VLA arc can be adequately
reproduced by judicious placement of sources.  Other features of the
VLA and Merlin maps are similar in that they can be reproduced
qualitatively by proper source placement in all our models, so they
do not
add information to our current modeling.  Ongoing
improvements in the radio maps, coupled with some type of CLEAN or
maximum entropy reconstruction algorithm for lenses
({\it e.g.\/} Wallington \etal\markcite{Wa1} 1996,
Ellithorpe \etal\markcite{El1} 1996) or other software for modeling
diffuse sources, will certainly be of use in testing and limiting the
lens models.

The NICMOS images are, at this writing, preliminary and may perhaps
yield quite useful constraints if sufficient S/N can be achieved.
Kochanek \etal\ (1998) state that the direction of extension
of the A and B arcs are sufficient to rule out the GN model.

\subsubsection{VLA Jets}
VLA images of the system (Greenfield \etal\markcite{Gr2} 1985)
show extended jets (labelled C, D, and
E) extending several arcseconds from the A quasar.  The absence of
jet images about the B quasar could be used as a model constraint. In
all of our models, the source regions for the C, D, and E jets 
are either singly imaged, or perhaps have highly demagnified images
leading from B toward the center of G1.  Avruch \etal\ (1997) may have
detected a counterimage of a low-surface-brightness extension of the E
jet, but the resolution is as yet too poor to yield much information.

\subsubsection{No Quasar C}
The failure to detect a third quasar image is not a useful constraint
on our models.  As noted in Paper II, the stellar light density
of G1 continues a power-law increase toward the center, as do all
other observed elliptical galaxies (Gebhardt \etal\markcite{Ge1}
1996). For surface-density power laws with exponents near the
isothermal value of $-1$, the third image will be absent or highly
demagnified.  The third image 
is also easily ``captured'' by a star and further demagnified.  The
properties of the third image might constrain the mass distribution in
the central 0\farcs1 of G1, but this mass has little effect on \hn.

\subsubsection{G1 Velocity Dispersion}
\label{tonry}
High-accuracy measurements of the stellar velocity dispersion of G1
are presented by Tonry \& Franx\markcite{To1} (1998; see also
Falco \etal\markcite{Fa2} 1997).  While the
velocity dispersion of G1 may be used to break the mass-sheet
degeneracy (\S\ref{weak}), we mention here a different use.  Over a
range of $\pm3\arcsec$ from the center of G1, Tonry \& Franx detect a
change of $\lesssim10\%$ in the velocity dispersion.  Given their good
seeing (0\farcs7) and the sharp central cusp in the G1 luminosity
profile (Paper II), this measurement shows that the velocity
dispersion of G1 is nearly constant over a range 1\arcsec--3\arcsec\ in
projected radius.  Thus the radial mass profile of G1 is nearly isothermal,
$\alpha=-1$ in the notation of Equation~(\ref{ellip1}) below.  A full
consideration of the constraints imposed by this measurement is beyond
the scope of this paper---see Romanowsky \& Kochanek\markcite{Ro2}
(1998) for the significant steps toward constraining the G1 mass
with stellar velocity data.  We will, however,  assume the very crude and
conservative constraint that the power-law index of the
projected G1 mass density at $\approx1\arcsec$ radius satisfies
$-1.5\le\alpha\le-0.5$.  
In a naive interpretation ({\it i.e.\/} spherically
symmetric galaxy with isotropic velocities), a mass index at
these upper or lower limits
would lead to a 30\% rise or fall (100~km~s$^{-1}$) in velocity
dispersion in the data of Tonry \& Franx, which can clearly be excluded.

\subsubsection{Cluster Location}
\label{ccenter}
The weak lensing mass map in Paper I shows the peak of the cluster
mass distribution located to the northeast of G1.  This displacement
is only marginally significant ($\approx1.5\sigma$); one cannot from
this data alone exclude the possibility that the cluster is centered
on G1.  Paper I also shows that the light distribution of the cluster
galaxies is peaked to the NE of G1, though again not at high
significance.  In what follows we will find further
evidence from the strong-lensing models that the cluster
mass peaks in the NE or SW quadrant.  We believe that the concurrence of
these three weak, but independent, lines of evidence is
sufficient to apply a constraint that the cluster mass density be
increasing to the NE quadrant of G1.

\section{Mass Models}
\label{models}

\subsection{Terminology}
We define in the usual fashion the dimensionless 2d gravitational
potential $\phi(\vec x)$ via
\begin{equation}
\nabla^2\phi(\vec x) = 2 \kappa(\vec x) = 2 \Sigma(\vec x) / \Sigma_{\rm
crit},
\end{equation}
where the critical density $\Sigma_{\rm crit}$ is determined by the
angular diameter distances $D_{OL}$, $D_{LS}$, $D_{OS}$ between
observer, lens, and source:
\begin{equation} 
\label{scrit}
\Sigma_{\rm crit} =
{ {4\pi c^2} \over G } { {D_{OS}} \over D_{OL}D_{LS}}.
\end{equation}
For our application the redshift of the observer is zero, of the lens
is $z_L=0.356$ (Tonry \& Franx 1998), 
and of the sources $z_S=1.41$ (Weymann \etal\markcite{We1} 1979).  The
position $(u,v)$ in the source plane of an object viewed at $(x,y)$ in
the image plane is
\begin{equation}
\begin{array}{lcr}
 u & = & x - \phi_{,x} \\
 v & = & y - \phi_{,y} 
\end{array}.
\end{equation}
Subscripts after the comma denote differentiation with respect to the
given coordinate(s). The inverse of the magnification matrix in this
region is 
\begin{equation}
\label{invmag}
\bf M^{-1} = \left( 
\begin{array}{cc}
1 - \phi_{,xx} &
-\phi_{,xy} \\
-\phi_{,xy} &
1 - \phi_{,yy} \\
\end{array}
\right)
\end{equation}

The time delay between two images at $\vec x_A$ and $\vec x_B$  of the
same source at $\vec u$ is
\begin{equation}
\label{dt1}
\Delta t_{AB} = (1+z_L){ {D_{OL}D_{OS}} \over {c D_{LS}} }
	\left[ { 1 \over 2} \left( |\vec x_A - \vec u|^2 - 
	|\vec x_B - \vec u|^2\right)
	- \phi(\vec x_A) + \phi(\vec x_B) \right]
\end{equation}
The mass sheet degeneracy is as follows:  if $\phi$ is a potential
which satisfies all lensing constraints with source positions $\vec
u_i$ for the various lensed objects and a time delay $\Delta t$, then
the alternate potential
\begin{equation}
\label{kap}
\tilde\phi(\vec x) = { {\kappa_c} \over 2} |\vec x|^2  + (1-\kappa_c)
\phi(\vec x)
\end{equation}
will satisfy all lensing constraints if sources are placed at
$(1-\kappa_c)\vec u_i$.   The first term of the equation is the potential
produced by a mass sheet of constant density $\kappa_c$ (in units of
the critical density).  Since the source plane is unobservable this
solution cannot be distinguished from the original in
Equation~(\ref{dt1}).
The time delay becomes
\begin{equation}
\label{dt2}
\Delta t_{AB} = (1-\kappa_c)(1+z_L){ {D_{OL}D_{OS}} \over {c D_{LS}} }
	\left[ { 1 \over 2} ( |\vec x_A - \vec u|^2 - 
	|\vec x_B - \vec u|^2)
	- \phi(\vec x_A) + \phi(\vec x_B) \right]
\end{equation}
For a given measured time delay, the distance scale (\hn) must change by
$1-\kappa_c$. 

In \S\ref{weak} we will break the mass-sheet degeneracy using the
ability of weak lensing measurements to determine the mean surface
mass density $\bar\kappa_R$ of mass within a circle of radius $R$.
We will take this circle to be centered on G1.  Let the original
potential $\phi$ have a mean mass density $\bar\kappa_{R,0}$
within radius R of G1.  For the altered potential $\tilde\phi$, the
weak lensing aperture mass measurement will give
\begin{eqnarray}
\bar\kappa_R  & = & \kappa_c + (1-\kappa_c) \bar\kappa_{R,0} \nonumber\\
\label{kaps}
\Rightarrow \;
(1-\kappa_c) & = & { {1-\bar\kappa_R} \over {1-\bar\kappa_{R,0}} } 
\end{eqnarray}
Combining Equations~(\ref{dt2}) and (\ref{kaps}) gives 
\begin{equation}
\label{dt3}
\begin{array}{ccl}
\Delta t_{AB} & = & \displaystyle
	(1+z_L){ {D_{OL}D_{OS}} \over {c D_{LS}} }
	(1-\bar\kappa_R) \Delta\hat t \\[12pt]
\Delta\hat t & \equiv & \displaystyle
{{ { 1 \over 2} \left( |\vec x_A - \vec u|^2 - 
	|\vec x_B - \vec u|^2\right)
	- \phi(\vec x_A) + \phi(\vec x_B) } \over 
	{1-\bar\kappa_{R,0}} }. \\
\end{array}
\end{equation}
The dimensionless quantity $\Delta\hat t$ depends only upon the
original lens model $\phi$.  The mass sheet degeneracy is broken by
measuring $\bar\kappa_R$ with weak lensing.  
Standard formulae for the angular diameter distances then give
\begin{equation}
\label{dt4}
\begin{array}{rcl}
H_0 & = & \displaystyle
	{ { 0.475 } \over {\Delta t_{AB}} } f(\Omega,\Lambda) (1-\bar\kappa_R) 
	\Delta\hat t, \\[12pt]
 &  = & (9.56\pm0.07\, {\rm km\,s^{-1}\,Mpc^{-1}}) f(\Omega,\Lambda) 
	(1-\bar\kappa_R) 
	\Delta\hat t. \\
\end{array}
\end{equation}
The numerical prefactor
assumes that $\vec x$ has units of arcseconds.  We take
$\Delta t_{AB} = 417\pm3$~days as determined by Kundi\'c \etal\
(1996) and confirmed by Haarsma \etal\ (1999).  We 
assume $\Omega=1$, $\Lambda=0$, and the filled-beam
approximation in calculating the angular diameter distances.  The
function $f(\Omega,\Lambda)$ expresses any further dependences on the
cosmic geometry, with $f(\Omega=1,\Lambda=0)\equiv1$.  For reasonable
cosmologies, $f$ varies by $\lesssim5\%$.  As our final \hn\ value is
uncertain by $\pm25\%$, we will henceforth ignore the corrections for
departures from the Einstein-de~Sitter geometry.

\subsection{Motivations for Mass Models}
\subsubsection{The Prime Directive}
The goal of this work is to determine \hn.  From this point of view,
{\it any} model mass distribution which is astrophysically reasonable
and can reproduce the observed strong lensing geometry to within
measurement errors must be considered a viable model.  Though a
simple class of models may provide a good fit to the observed
lensing geometry, we are obliged to investigate whether
added complexity will extend the range of
permissible \hn\ values.  Previous models of this lens have considered
the galaxy to have a power-law mass profile, sometimes with elliptical
shape and/or a softened core or central point mass.  There are {\it
no} galaxies for which the global mass distribution is known to 
fit any simple parameterization to the $\sim1$\%
accuracy required to fit the lensing constraints in this
system.  It behooves us, therefore, to give our model galaxy the
freedom to depart from an elliptical power law, and see whether this
allows a wider range of \hn\ or permits a better fit.  Keeping in mind
that our knowledge of dark matter distributions in galaxies and
clusters is sketchy at best, we propose some desirable generalizations
of previous models here.

\subsubsection{Break the Power Law}
\label{circsym}
Consider a very simple doubly imaged system in which the potential is
circularly symmetric, and our two quasar images appear astride the
lens center at radii $r_A$ and $r_B$.  
To obtain \hn\ we need the difference in
potentials $\phi(r_A)-\phi(r_B)$.  The requirement that images A and B
have a common source determines the derivative
$\phi^\prime(r_A)+\phi^\prime(r_B)$, and flux ratio between A and B
constrains the second derivatives $\phi^{\prime\prime}(r_A)$ and
$\phi^{\prime\prime}(r_B)$.  The heart of the problem in determining
\hn\ is that the lensing optics constrain the derivative(s) of
$\phi$, not $\phi$ itself as needed for \hn.
The potential ($\phi$) difference between A and B
is equal to a line integral of the deflection ($\phi^\prime$)
on any path from A to B.  So to constrain \hn\
accurately we need to measure the deflection at radii intermediate to
$r_A$ and $r_B$, or we must make assumptions about the behavior of the
potential at these intermediate radii.

A power-law model $\phi(r)=br^{\alpha}$ has two parameters, and hence
the entire potential is specified by the positions
and flux ratio of the A and B images.  In our case $r_A$ and $r_B$
are  5.22\arcsec\ and 1.03\arcsec, respectively, so the power-law
assumption amounts to integrating $\phi^\prime$ across a factor of 5 in
radius based on our constraint of $\phi^\prime$ and $\phi^{\prime\prime}$
at these endpoints.
To allow the widest range in \hn, we should permit $\phi^\prime$ more freedom
between A and B.  We will implement this by
investigating models in which the power law breaks at some position
between $r_A$ and $r_B$ (adding 2 degrees of freedom to even the
simple circularly symmetric case).  A glance forward to
Figure~\ref{profiles} shows that these simplified assertions on freedom
in the radial profiles are borne out by the detailed modelling.

The WFPC2 objects (Paper II) should be of great help in constraining
\hn\ because they fall at several different radii between $r_A$ and $r_B$, and
hence help ``tie down'' the behavior of $\phi^\prime$ between the two quasars.

\subsubsection{Twisting Mass Contours}
The $0957+561$ lens is clearly not circularly symmetric since QA, QB,
and G1 are not colinear.  The light distribution is elliptical, and an
elliptical mass distribution should be expected as well.  Even a
modest ellipticity of 10\% can change the mass density, and
hence the magnification, at the position of a quasar by $\sim10\%$,
well above the measurement error.  The isophotes of G1 are observed to
twist by 10\arcdeg\ or so between $r_A$ and $r_B$ (Paper II), so we
should investigate the possibility that the mass contours do likewise.  Since
this twist could alter the A/B magnification ratio substantially, we should
investigate possible effects on \hn.

\subsubsection{Higher-Order Cluster Approximation}
Kochanek
\markcite{Ko1} (1991) investigates the degeneracies of $0957+561$
lens models given the VLBI data, and demonstrates that the models are
highly underconstrained.  A very large range of
astrophysically plausible mass distributions can reproduce the
geometry of the quasar and jet images, {\it even when the galaxy
cluster mass density near the strong-lensing region is held fixed.\/}
Most other models (FGS, GN) approximate the effect of the lensing
cluster to quadratic order in an expansion about G1:
\begin{equation}
\label{clust1}
\phi_c \approx \phi_0 + \phi_{c,x} x + \phi_{c,y} y
 + \kappa_c(x^2+y^2)/2 + \gamma_+(x^2-y^2)/2
 + \gamma_\times xy.
\end{equation}
The first three terms have no observable consequences and may be
ignored; the $\kappa_c$ term is the degenerate mass sheet term
discussed earlier, and the last two terms give a constant shear
specified by the two parameters $(\gamma_+,\gamma_\times)$.
Kochanek shows that {\it cubic}-order terms in the power-law expansion
of the cluster potential give significant deflections in the
$0957+561$ system for typical expected softened-isothermal cluster
mass distributions.  We will test the effect of higher-order cluster
terms upon the model fits, and in fact show that an adequate fit is
now attainable only if such terms are included in a model.

Because the shape of the cluster mass distribution cannot be tightly
constrained with weak lensing measurements (Paper I), we must also
allow for the possibility of substructure or other departures from the
isothermal profiles often assumed.  Our philosophy will be to assume
only that the cluster potential is ``smooth'' over the strong-lensing
region ({\it i.e.\/} within 30\arcsec\ of the G1 center) in the sense
that the importance of terms in the power-law expansion 
decreases with increasing order.  We will {\it not} force
$\kappa_c$, $\gamma_\times$, $\gamma_+$, and the higher-order
power-law coefficients to have the relative values required for an
isothermal cluster.  The independence of these coefficients in our
models means that the cluster is allowed to be asymmetric or lumpy.

\subsection{G1 Dark Matter Models}
All of our models for the G1 mass distribution build the mass as a sum
of one or more elliptical power-law distributions over circular
annuli.  In polar coordinates centered on G1, the surface density for
the $i^{\rm th}$ term is
\begin{equation}
\label{ellip1}
\begin{array}{cc}
{ {\Sigma_i(r,\theta)} \over {\Sigma_{\rm crit}} } =
\left\lbrace \begin{array}{cc}
  0 & r<r_{{\rm inner},i} \\
  b_i s_i^{\alpha_i} & r_{{\rm inner},i} < r <r_{{\rm outer},i} \\
  0 & r_{{\rm outer},i}>r \\
\end{array} \right.,
\\
  s_i^2 = r^2 \left[1-e_i\cos 2(\theta-P\!A_i)\right] (1-e_i^2)^{-1/2}.
\end{array}
\end{equation}
Contours of constant $s_i$ are ellipses with ellipticity $e_i$ and
major-axis position angles $P\!A_i$.  $\alpha_i$ and $b_i$ specify the
radial profile and normalization of the mass distribution.
The potential and its first and second derivatives can be
accurately calculated with compact analytic formulae to arbitrary
precision.  The formulae for the multipole expansions of these
elliptical mass distributions are given in Appendix~\ref{mpole}.

Our first {\it a priori} constraint is that $e\le0.6$ (axis ratio
less than 2:1).  The isophotes of G1 reach $e\approx0.4$ at a
radius of 20\arcsec, and a matter distribution significantly flatter
than 2:1 would be difficult to believe.  We also enforce
$-1.95<\alpha<-0.05$, to avoid divergences at large or small radii.
As mentioned in \S\ref{tonry}, we will require $-1.5\le\alpha\le-0.5$ for the
dominant mass component at $r\approx1\arcsec$, since the dynamical
evidence for G1 and most other elliptical galaxies suggests
mass profiles near isothermal.

We now describe several parameterizations of the dark matter in G1
that we have tried in lens models.  Each is assigned a short code
(given in the section headings) for a compact designation of models.

\subsubsection{DM1: Single-Zone Model (4 Parameters)}
The baseline model for the G1 mass distribution is a single power-law
ellipse with $\alpha$, $b$, $e$, and $P\!A$ free to vary.  We take
$r_{\rm inner}=0$ and $r_{\rm outer}=30\arcsec$, spanning the entire
strong-lensing area.

\subsubsection{CORE: Softened Single-Zone Model (5 Parameters)}
Many models ({\it e.g.\/} FGS, GN, BLFGKS) allow the G1 dark matter to flatten
inside some ``core radius'' $r_c$.  We can mimic this behavior
with the power-law formalism with surface density
\begin{equation}
{ {\Sigma(r,\theta)} \over {\Sigma_{\rm crit}} } =
\left\lbrace \begin{array}{cc}
  b r_c^{\alpha} \left[ (1-\alpha/2) + \alpha/2 (s/r_c)^2 \right ]
  & 0<r<r_c \\
  bs^{\alpha} & r_c < r < 30\arcsec \\
\end{array} \right.
\end{equation}
The quadratic profile inside $r_c$ reaches zero slope at the origin
and matches the level and slope of the power law outside $r_c$.  The
free parameters are those of the DM1 model plus the core radius $r_c$.

Note that the annular multipole method produces {\it elliptical} mass
distributions (functions of $s$) bounded by {\it circular} limits
(bounds in $r$).  For $e\ne0$, the value of $s$
varies over some finite range as we travel around the circle 
at $r=r_c$.  The quadratic and the power-law do not match exactly
over this finite range in $s$, so there can be discontinuities
in $\Sigma$ at the $r_c$ circle.  For nearly-isothermal CORE models,
the fractional jump in $\Sigma$ at the boundary is
$\approx{3\over2}e^2$.
The lens potential and deflection
are continuous across this boundary but the magnification is not.  For
this reason, we make sure that $r_c$ does not lie in the
$\sim50$~mas space between quasar B
and jet B5, or between quasar A and jet A5.

\subsubsection{DM2: Two-Zone Model (8 Parameters)}
A more complex G1 mass distribution allows for the possibility of
breaks in the power law and ``isophotal'' twist.  This is implemented
by a two-zone galaxy mass model:
\begin{equation}
\label{dm2}
{ {\Sigma(r,\theta)} \over {\Sigma_{\rm crit}} } =
\left\lbrace \begin{array}{rl}

  bs_0^{\alpha_0} & 0< r < r_{01} \\
  br_{01}^{\alpha_0-\alpha_1}s_1^{\alpha_1} & r_{01} < r < 30\arcsec \\
\end{array} \right. .
\end{equation}
The free parameters are $\lbrace b, \alpha_0, \alpha_1, r_{01},
e_0, e_1, P\!A_0, P\!A_1 \rbrace$.  To keep the mass distribution
reasonable we limit the twist to $|P\!A_0-P\!A_1|\le10\arcdeg$ and the
ellipticity change to $|e_0-e_1|<0.2$.  To reduce the number of free
parameters we can enforce $e_0=e_1$, $P\!A_0=P\!A_1$, and/or
$\alpha_0=\alpha_1$.  There will again be a density
discontinuity at $r_{01}$, of fractional strength 
$\approx (e_0\alpha_0-e_1\alpha_1)\lesssim20\%$.  Because of
the resultant magnification discontinuity, we will keep the join
radius away from the quasar jets.

\subsubsection{DM3: Three-Zone Model (12 Parameters)}
A more complex model allows three power-law zones, joined at radii
$r_{01}$ and $r_{12}$.  The equations for $\Sigma$ are analogous to those
in Equation~(\ref{dm2}). The same {\it a priori} constraints are
applied to the change in $P\!A$ and $e$ at each joint.  This model has
as many parameters as we have constraints (not yet counting the
cluster parameters).

\subsection{Additional G1 Mass Components}
There are two other potentially significant mass components to G1
beyond the dark matter:  the luminous matter and a potential central
black hole.  Both components may be described using the elliptical
power-law formulae in Equation~(\ref{ellip1}).

\subsubsection{ML: Mass Traces Light (1 parameter)}
\label{mlsec}
The visible component of G1 is a significant deflector.  Dynamical
studies of nearby elliptical galaxies suggest that most of the matter
within $r_e$ is stellar. Our lens model should have $\Sigma(\vec x)$
be at least as large as the stellar component at all $\vec x$.  We can
enforce this by including a mass-traces-light term in $\Sigma$ and
requiring the $M/L$ ratio to be at least as large as the $M/L$ of the
stellar population of a giant elliptical galaxy.

Surface photometry of G1 is given in Paper II from the WFPC2 image and
the ground-based $R$-band image of Bernstein, Tyson, \&
Kochanek\markcite{Be2} (1993). The isophotes twist and the ellipticity
rises at outer radii, and a single power-law is a poor fit to the
radial profile.  We approximate this behavior by a two-zone power law
model:
\begin{equation}
\label{ml}
{ {\Sigma_\ast(r,\theta)} \over {\Sigma_{\rm crit}} } =
\left\lbrace
\begin{array}{llll}
0.16b_\ast s_{\ast 0}^{-1.3}, & e_{\ast 0}=0.15, & P\!A_{\ast 0}=141\arcdeg
 & {\rm for}\;0<r<2\arcsec \\
0.16b_\ast s_{\ast 1}^{-1.9} (2\arcsec)^{0.6}, 
	& e_{\ast 1}=0.3, & P\!A_{\ast 1}=146\arcdeg
 & {\rm for}\;2\arcsec<r<30\arcsec. \\
\end{array} \right.
\end{equation}
The ellipticity, $P\!A$, and mass profiles of this distribution pass
through nearly all the $1\sigma$ error bars of the surface brightness
profiles in Figure~2 of Paper II.  

We have chosen the prefactor in Equation~(\ref{ml}) so that $b_\ast=1$
gives the minimum mass density expected from the observed stellar
population of G1.  The derivation is given in Appendix~\ref{mlnorm}.
We enforce $b_\ast\ge1$ when we include the ML term; a higher value
might result if the Universe is large or open, or if some dark matter
traces the light.  We note here that models in which {\it all} the G1
mass traces the light give very poor fits to the lensing constraints,
and will not be considered further.

\subsubsection{BH: Central Black Hole (1 Parameter)}
Many previous models have allowed G1 to have a central massive black
hole.  We can include such a central mass as a term in
the form of Equation~(\ref{ellip1}).  We parameterize the central
black hole as having mass $M_9\times10^9\,h^{-1}M_\odot$.  
A central point mass in excess of $10^{10}\,M_\odot$ has never been
reliably detected in any galaxy; less reliable methods have suggested
central masses up to $3\times10^{10}\,M_\odot$, but only in the most
massive cD galaxies (Richstone \etal\markcite{Ri1} 1998).
The measured velocity dispersion of G1 (Tonry \&
Franx 1998) would not place it among these most massive galaxies, so we
enforce $0<M_9<10$ for the G1 BH term.  Even at
the upper mass limit, the central black hole has an Einstein radius
of only 0.3\arcsec; we will see that sensibly-sized black holes do not
have significant effects upon the model fits (see Figure~\ref{profiles}).

Some previous models ({\it e.g.\/} FGS) have derived BH masses 
of order $10^11\,h^{-1}M_\odot$ much larger than our limit.  
In these models, the ``black hole'' mass must represent a
concentration of mass in the central $r<1\arcsec$ of G1, not
just a true point mass.  Since our models allow for
the central concentration of stellar matter (ML term) or in the dark
matter (DM2 term), we will confine the $M_9$ parameter to the range
expected for an actual black hole.

\subsection{Cluster Models}
\subsubsection{C2: Quadratic Approximation (2 Parameters)}
The simplest treatment of the cluster mass distribution is the
quadratic approximation in Equation~(\ref{clust1}).  The constant and
linear terms in the expansion have no effect, and the $\kappa$ term
has no measurable effect on the strong-lensing region, so the only
free parameters are the shear coefficients $\gamma_+$ and
$\gamma_\times$.  The position angle of the shear major axis is
defined by $\gamma_\times / \gamma_+ = \tan2\theta_\gamma$.  If the
external shear is entirely due to a circularly symmetric cluster, then
$\theta_\gamma$ gives the PA of the cluster center with respect to the
G1 center.  If the cluster departs from circular symmetry or is not
the only source of shear along the line of sight, then $\theta_\gamma$
may not point to the cluster center.

\subsubsection{C3 (C3S): Cubic Approximation [6 (4) Parameters]}
Taking the expansion of the cluster potential to cubic order (and
dropping the constant, linear, and $\kappa$ terms) gives
\begin{equation}
\label{clust2}
\phi_c(r,\theta) = { 1 \over 2} \gamma r^2\cos2(\theta-\theta_\gamma)
 + {1 \over 4} \sigma r^3 \cos(\theta-\theta_\sigma)
 + {1 \over 6} \delta r^3 \cos 3(\theta-\theta_\delta).
\end{equation}
This notation is slightly changed from that of Kochanek (1991).  
It takes four additional parameters to specify the cubic-order terms.
The $\sigma$ term is the potential produced by the gradient of
the cluster mass near G1; $\sigma$ is the amplitude of the density
gradient (in critical units) and $\theta_\sigma$ gives the direction of
the gradient.  The $\delta$ term is the potential induced by the $m=3$
component of cluster mass exterior to G1.  It produces no convergence,
and a shear that varies linearly across G1.

If the cluster is circularly symmetric, then
$\theta_\gamma=\theta_\sigma=\theta_\delta$, all pointing toward the
cluster center.  Kochanek (1991) gives the relation between $\gamma$,
$\sigma$, and $\delta$ for the case in which the cluster has a
softened isothermal potential.  More generally we expect
$\sigma\approx\delta\approx\gamma^2\approx\kappa^2$ and a rough
agreement between the angles.  We will in most cases enforce
\begin{equation}
\label{c3sis}
\theta_\sigma=\theta_\delta \qquad {\rm and} \qquad \sigma=-2\delta/3
\end{equation}
(as for a
singular isothermal cluster) to reduce the number of free parameters,
which we will call the ``C3S'' cluster model.
We should keep in mind, however, that the cluster mass distribution could
easily be lumpy or asymmetric, so we cannot depend upon any tight
relations between these parameters.

\subsubsection{C4P: Cluster with Mass Peak at $R<30\arcsec$ (5
Parameters)}
\label{c4p}
The C3 cluster models assume that, inside our
R=30\arcsec\ canonical division between strong and weak lensing regimes,
the cluster mass can be described as a constant ($\kappa$ term) plus
linear gradient ($\sigma$ term).  The weak lensing data in Paper I 
give a marginally significant indication that the {\it total} mass density
continues to rise inside $R<30\arcsec$, which could be due in part to
a peak in the {\it cluster} mass density.  To accommodate this possibility we
can add a quadratic cluster-mass surface density to the model of the
form
\begin{equation}
\label{cr2}
{ {\Sigma(r)} \over {\Sigma_{\rm crit}} } =
\kappa_p\left[1-2(r/R)^2\right].
\end{equation}
The C3 cluster potential in Equation~(\ref{clust2}) represents a linear mass
gradient in direction $\theta_\sigma$, which can be combined with the
quadratic cluster density in Equation~(\ref{cr2}) to produce a
maximum in cluster mass density displaced from G1 in the direction
$\theta_\sigma$.  This allows us to model a cluster mass distribution
that reaches a quadratic maximum
anywhere within the $R=30\arcsec$ circle.  We
restrict $0<\kappa_p<0.5$, with the upper bound based upon the
weak-lensing analysis.  The cluster is in this case described by the
parameter set $\{\gamma,\theta_\gamma,\sigma,\theta_\sigma,\kappa_p\}$
if we enforce the ``C3S'' conditions in Equation~(\ref{c3sis}).
The potential contains a limited set of quartic terms.

Note that a very concentrated cluster mass peak, {\it e.g.} an
isothermal singularity, would be subsumed into the DM1 term if it were
centered on G1, and hence we do not need an additional ``cluster
mass'' term to allow for such behavior.

\subsubsection{C4S: Quartic Approximation (6 Parameters)}
\label{c4s}
To test the sensitivity of our results to yet higher-order elements of
the cluster potential, we can add quartic terms to
Equation~(\ref{clust2}). 
Including the fourth derivatives of the cluster potential with full
freedom would add 5 more parameters to the model.  A more manageable
approach, which still tests the importance of quartic terms, is to
require that the relative amplitudes of the quartic terms be those of
a singular isothermal sphere (SIS) 
cluster at some position angle $\theta_4$.  The
amplitudes of the quartic terms are scaled by a factor $c_4$.  The
free parameters for the quartic cluster are then
$\{\gamma,\theta_\gamma,\sigma,\theta_\sigma,c_4,\theta_4\}$.  The
quadratic, cubic, and quartic terms are not required to correspond to
the {\it same} SIS cluster direction or amplitude.

\section{Fits of Models to the Constraints}
\label{fits}
In this section we combine the various model mass components of the
previous section to the constraints of \S\ref{constraints}.  The
emphasis is on bounding the range of allowed $\Delta\hat t$.
We will
first describe our numerical methods, and then succeeding sections
will cover mass models with increasing complexity in their treatment
of the cluster mass.  The final part of this section is a summary and
discussion of the strong-lensing models.

\subsection{Numerical Methods}
\label{methods}
The figure of merit for fits to the constraints is the overall
$\chi^2$ for the hypothesis that there is a single source for
each of the 4 pairs of
images---quasar cores A \& B; jets A5 \& B5; Blobs 2 \& 3; and Knots 1
\& 2.  The total $\chi^2$ is defined as
\begin{equation}
\label{chi2a}
\chi^2 = \sum_{\rm pairs} (\chi^2_{\rm posn} + \chi^2_{\rm flux}).
\end{equation}
The sum is over the 4 image pairs.  For a given pair, consisting of an
image A and an image B, the flux ratio $\chi^2$ is straightforward as
\begin{eqnarray}
\label{chi2b}
\chi^2_{\rm flux} & = & \left\lbrace
\begin{array}{l}
\left[ (f_B/f_A)_{\rm model} - (f_B/f_A)_{\rm meas}\right]^2/\sigma^2
\\
\multicolumn{1}{c}{\rm or} \\
\left[ \ln(f_B/f_A)_{\rm model} - \ln(f_B/f_A)_{\rm
meas}\right]^2/\sigma^2
\end{array}
\right. \\
(f_B/f_A)_{\rm model} & = & |\det {\bf M^{-1}_A} | / |\det {\bf
M^{-1}_B} |,
\end{eqnarray}
where the inverse magnification matrix is given in
Equation~(\ref{invmag}), and the predicted flux ratio and its
uncertainty $\sigma$ (either in linear or in log space) are listed in
Table~\ref{tconstr}.  The linear form of the error is used for the
precisely known quasar and jet flux ratios, while the logarithmic form
is applied to the poorly know arc and blob flux ratios.

The positional $\chi^2$ is more complex.  First, we take the image
positions ${\bf x}_A$ and ${\bf x}_B$ and map them back to source
plane positions ${\bf u}_A$ and ${\bf u}_B$.  We must also map the
observational error ellipse for each image back into the source
plane.  Let the variance matrix for ${\bf x}_A$ be called
${\bf \Sigma}_{A,i}$ (the $i$ denotes image plane).  Under the
assumption that the lens map is linear across the error ellipse, the
variance matrix for the {\it source} position of image A becomes
\begin{equation}
\label{chi2c}
{\bf \Sigma}_{A,s} = {\bf M^{-1}_A} \, {\bf \Sigma}_{A,i} \, {\bf
M^{-1}_A}
\end{equation}
The $\chi^2$ for the hypothesis that images A and B have a common
source position can then be expressed as
\begin{equation}
\label{chi2d}
\chi^2_{\rm posn} = \left({\bf u}_A-{\bf u}_B\right)^T
	\, \left({\bf \Sigma}_{A,s}+{\bf \Sigma}_{B,s}\right)^{-1}
	\, \left({\bf u}_A-{\bf u}_B\right).
\end{equation}
Near the caustics the assumption of a locally linear mapping
may fail, but the alternative is to set the source positions as
free parameters in the model, which would greatly increase the number
of dimensions we would have to search for our extrema.

For each of the mass models discussed below, we first search to
minimize $\chi^2$ over the parameter space.  The model is considered
to be viable if it produces a $\chi^2$ with $\nu$ degrees of freedom
such that the probability $Q(\chi^2,\nu)$ of exceeding the given
$\chi^2$ purely due to the observational errors is at least
5\%.  If a model can meet this test, we then proceed to find the
parameter values that produce the minimum and maximum $\Delta\hat t$
values subject to the constraint that $Q(\chi^2,\nu)\ge0.05$.  This we
assign as the 95\% confidence interval for $\Delta\hat t$ for the
model. Note that this differs from the usual $\Delta\chi^2$ method,
which is used to place confidence limits on fitted parameters when
one is sure that the parametric model is correct.  The $\Delta\chi^2$
method is not easily applied to our situation, in which we are
considering a range of models with different degrees of freedom, and
will accept any model that is consistent with the observations.

The models have up to 12 free parameters, which we
want to optimize for either the lowest $\chi^2$, or for the
highest/lowest $\Delta\hat t$ subject to a maximum on $\chi^2$.  We
have adopted a few strategies to make these multi-dimensional searches
faster and more likely to find the appropriate global minimum.  

All of the mass distributions discussed in \S\ref{models} have
analytic expressions for the potential $\phi$ and its derivatives.
For the elliptical mass distributions, the potential is expanded in
multipoles as described in Appendix~\ref{mpole}.  The coefficients of
the multipoles are expressed as power series in the ellipticity $e$.
We retain terms sufficient to approximate the
elliptical mass distribution to accuracy of 1\% or better for the
ranges of $e$ and $\alpha$ we allow.

From Equations~(\ref{chi2a})--(\ref{chi2b}) we see that $\chi^2$
depends upon the model through the source positions and magnification
matrices at each image position, which are in turn specified by the first
and second derivatives of the potential $\phi$ at each of the 8 image
locations listed in Table~\ref{tconstr}. 
Evaluating $\Delta\hat t$ for a model requires additionally
that we calculate the potential $\phi$ at the two quasars, and the
mean density $\bar\kappa_{R,0}$.

To reduce the dimensionality of the parameter search, we exploit the
linear methods of Kochanek (1991).  The potential $\phi$ and its
derivatives are linear in many of the lens parameters, such as the G1
normalization $b$ and the shear parameters $\gamma_+$ and
$\gamma_\times$.  The equations expressing the constraint that two
images have a common source are linear in the derivatives of $\phi$,
and hence in these linear model parameters as well.  Thus any {\it
exact} constraint on multiple images can be used to eliminate lens
parameters.  We consider the positions of quasars A and B to be known
exactly, and use this to express $\gamma_+$ and $\gamma_\times$ in
terms of $b$ (given values of the other model parameters).  The
expressions above for $\chi^2$ can then also be reduced to
ratios of polynomials in $b$, which are messy but rapidly calculable.
The scaled timed delay $\Delta\hat t$ is a ratio
of two linear functions of $b$.  Given the values of the other
parameters, we can therefore optimize $\chi^2$ or $\Delta\hat t$ over
the parameters $b$, $\gamma_+$ and $\gamma_\times$ using rapid
one-dimensional search methods.

The search over the remaining linear and non-linear parameters is done
using the Adaptive Simulated Annealing code\footnote{
ASA software available at http://www.ingber.com, developed by Lester
Ingber and other contributors.}
(``ASA'', Ingber\markcite{In1},
1996).  Simulated annealing is particularly useful for finding the
global minimum of a function which may have many local minima.
For a given model we repeat the ASA search twenty or so times with
different random number seeds in a further effort to sample the full
phase space for global minima.  With more than a few dimensions to
search, ASA can become slow to converge on the bottom of a given
``valley'' in the merit function.  We therefore use the output of ASA
as the starting point for an optimization using the downhill-simplex
program SUBPLEX (Rowan\markcite{Ro3} 1990).  
Thus the simulated annealing is used
in a global search for regions of low $\chi^2$, and the downhill
simplex is used to ``tune up'' the fit within these regions.
It typically takes an hour to complete 20 optimizations of a
given model on a 200-MHz Pentium processor running Linux.

\subsection{Models with Second-Order Cluster}
The simplest models to fit to the lens constraints use the ``C2''
cluster approximation of a convergence and shear.  The models of
FGS, GN, and 4 of the 5 models of BLFGKS use this approximation.  
In short, we find that {\it
no models with second-order cluster match the observations.\/}  All
have $\chi^2/\nu$ values outside the 95\% CL bounds.  This situation
remains true even when we allow the G1 mass distribution to have
considerable freedom.  The following paragraphs provide
more detail on the results of our fits to C2 models, and some
comparison with models by other authors.  Table~\ref{tc2} lists the parameters
for the best-fitting models of each type.

\begin{table}
\epsscale{1.0}
\plotone{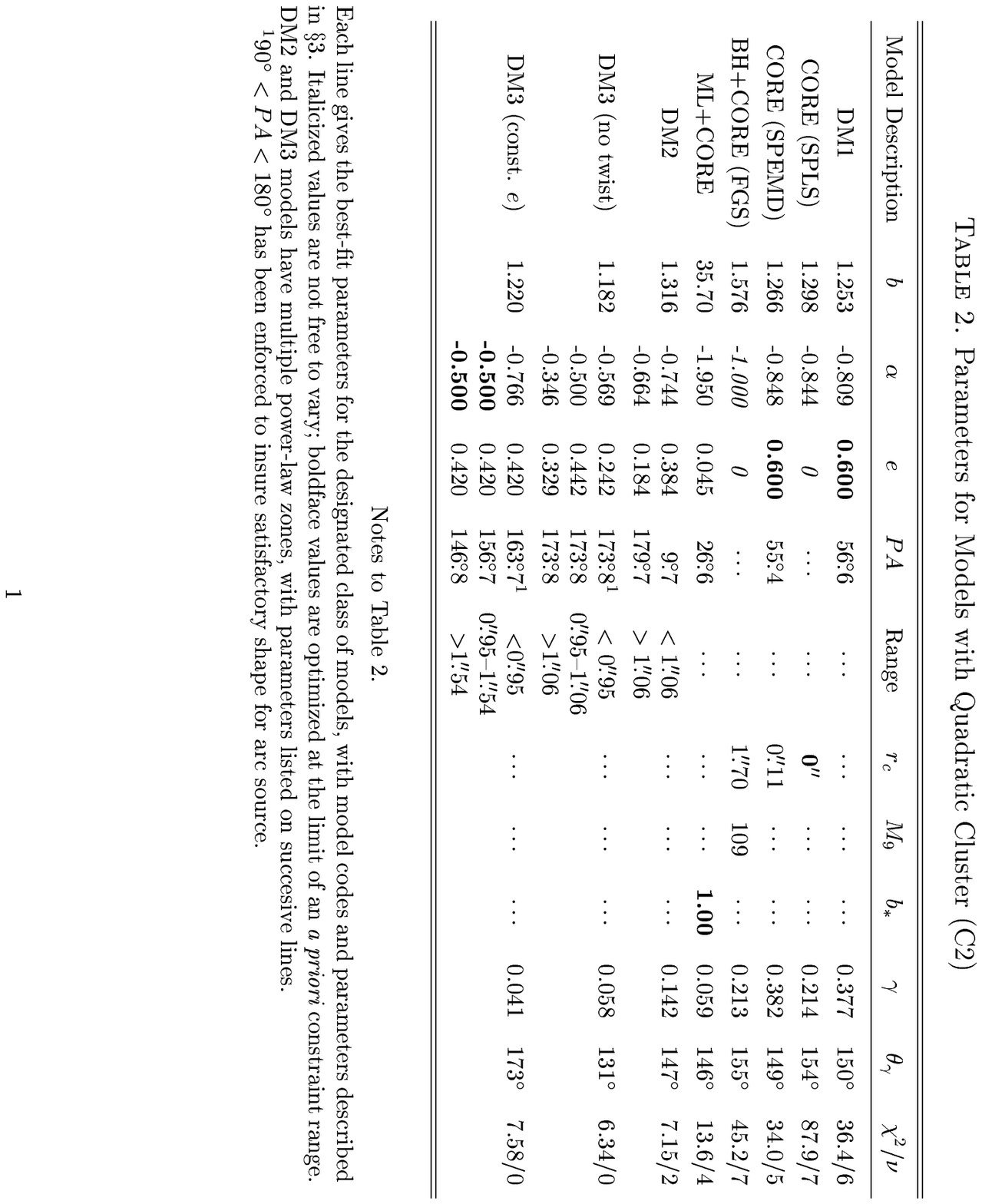}
\label{tc2}
\end{table}

\subsubsection{Baseline Model: C2+DM1 (6 DOF)}
Our simplest model uses the C2 cluster and makes G1 a single-zone elliptical
power-law mass distribution (DM1).  The best-fit parameters are listed
in Table~\ref{tc2}.  These parameters yield $\chi^2/\nu=36.4/6$, which would
occur by chance only $Q=2\times10^{-6}$ of the time, so the model is
strongly excluded.  The model is further excluded because it does not
satisfy our qualitative constraint of collapsing the arc image into a
compact source---this is a consequence of the G1 alignment
$PA=56\arcdeg$, which is nearly orthogonal to the observed light
distribution.  Constraining the G1 $PA$ to be nearer the visible $PA$
does collapse the arc but raises the overall $\chi^2$ to 55.
In either case G1 is highly elliptical and slightly shallower than
isothermal for the best fit.

\subsubsection{Comparisons with Previous Models:  C2+CORE, C2+BH+CORE
(5--7 DOF)}
Adding a core radius to G1 (mass distribution ``CORE'') and/or a black
hole (``BH'') provides a means to mimic the mass models of FGS,
GN and BLFGKS.  The comparison is not exact because our constraints
are different from these authors' as is our analytic formulation of
the softened-core mass distribution.  We find that our best-fit
parameters and time delays agree well with comparable models by other
authors, despite differences in these details.

The ``SPLS'' model of GN (and of BLFGKS) has a circular power-law
galaxy with softened core and is similar to our model C2+CORE with the
restriction $e=0$ (rendering the $PA$ irrelevant).  Our best fit to
this model has $\chi^2/\nu=87.9/7$.  Despite the slightly different
formulations, we obtain very similar results to GN and BLFGKS, in that
all three best-fit models require:  G1 slightly shallower than
isothermal, with $\alpha=-0.84$; core radius optimized at zero; and
external shear of $\gamma\approx0.22$ oriented with
$\theta_\gamma\approx155\arcdeg$ (using our notation).  Of course all
authors agree that the reduced $\chi^2$ of this model is well above
unity.  The $\Delta \hat t$ values agree within 6\% among the three
fits to this model.

The ``SPEMD'' model of BLFGKS is an {\it elliptical} softened
power-law with quadratic cluster, and hence is conceptually similar to
our model C2+CORE.  Our best-fit parameters yield
$\chi^2/\nu=34.0/5$.  Our model, like that of BLFGKS, makes G1 highly
elliptical (pushed to the $e=0.6$ {\it a priori} bound in our case) at
$PA=56\arcdeg$, orthogonal to the visible light, with core radius
$\approx0\farcs1$.  The BLFGKS galaxy is nearly isothermal whereas we
fit $\alpha=-0.85$, slightly shallower.

The FGS model (also re-fit by GN and BLFGKS) is conceptually similar
to our C2+BH+CORE model, restricted to $e=0$ and $\alpha=-1$. Our
best-fit model gives $\chi^2/\nu=45.2/7$, again agreeing with GN and
BLFGKS on the poor fit to the data.  Our parameters are quite close to
those of the FGS and the
FGS-like solution in BLFGKS, demanding a black hole mass
of $1.1\times10^{11} h^{-1} M_\odot$ (which we would normally exclude
as astrophysically implausible)
and an external shear of
$\gamma=0.21$ at $\theta_\gamma=155\arcdeg$. 

\subsubsection{Additional Components: BH, ML (4--5 DOF)}
Addition of a central black hole of reasonable mass ($\le10^{10}h^{-1}
M_\odot$) to the baseline model
does not improve the fit.  The best-fit parameters for C2+BH+DM1
models set the black hole mass to zero, and give the same $\chi^2$ as
the baseline C2+DM1 model.

An astrophysically attractive model is C2+ML+CORE, in which G1 
has a central mass cusp due to stars, and a softened dark-matter
halo is added.  This family yields a best-fit model with
$\chi^2/\nu=13.7/4$, having only $Q=0.008$ chance of being consistent
with the constraints.

\subsubsection{Two-zone Galaxy:  C2+DM2 (2 DOF)}
The model C2+DM2 gives G1 the freedom to have a break in its power
law, isophotal twist, and a change in ellipticity.  The best-fit
parameters give $\chi^2/\nu=2.27/2$ ($Q=0.32$), thus this model is
fully consistent with the imaging constraints.  It is not, however,
an astrophysically plausible model:  the index of the projected mass
power law is $\alpha_0=-0.35$ within $r_{01}=0\farcs78$ and an even
shallower $\alpha_1=-0.07$ outside the join radius.  The stellar
dynamical measurements discussed in \S\ref{tonry} would certainly
prove this model to be astrophysically implausible or impossible.

To produce more plausible parameters for the C2+DM2 model, we
constrain $-1.5\le\alpha_1\le-0.5$.  The best-fit model now has
$\chi^2/\nu=7.14/2$ ($Q=0.028$), no longer an adequate fit to the
observations.  Unlike the shallow-mass model in the previous
paragraph, this model produces $\Delta\hat t$ consistent with the
better-fitting models discussed below.

\subsubsection{Three-zone Galaxy: C2+DM3 (0 DOF)}
The C2+DM3 model gives the G1 mass even more flexibility, but has 14
free parameters for the 12 observational constraints.  In an attempt
to avoid astrophysically implausible solutions, we demand that the G1
mass profiles be either convex ($\alpha_0\ge\alpha_1\ge\alpha_2$) or
concave ($\alpha_0\le\alpha_1\le\alpha_2$) in log-log space. To reduce
the number of free parameters, we investigate two restricted forms of
the C2+DM3 models:  first, the ``no twist'' models, which have the
restriction $PA_0=PA_1=PA_2$, and second, the ``constant $e$'' models,
which are restricted to $e_0=e_1=e_2$.  In other words we let either
the $PA$ or ellipticity vary with radius, but not both.  Each subclass
of models has 12 free parameters, meaning that there are formally zero
degrees of freedom.  We will judge these models as if they have
$\nu=1$ degree of freedom.

The best-fitting ``no twist'' model has $\chi^2/\nu=4.17/1$
($Q=0.041$), marginally excluded. But this model also has the G1 mass
oriented at $PA=22\arcdeg$, which leads to an arc source that is
unacceptably extended.  Constraining $90\arcdeg < PA < 180\arcdeg$
gives a best-fit model at $PA=180\arcdeg$ which has a marginally
acceptable arc source.  This model yields $\chi^2/\nu=6.34/1$
($Q=0.011$), which is excluded.

The best-fitting ``constant $e$'' model again has galaxy mass oriented
perpendicular to the light and does not collapse the arc.  With the
above restriction on galaxy orientation implemented, the best-fit
parameters now orient the G1 mass with the light and give a good arc
source.  This model yields $\chi^2/\nu=7.58/1$
($Q=0.006$), which is excluded.

\subsection{Models with Cubic-Order Cluster}
With the addition of cubic-order cluster terms to our models we find
that even our simplest G1 mass models produce acceptable fits to the
constraints.  Once such an acceptable fit is identified, we turn our
attention to exploring the range of \hn\ spanned by acceptable
models.  We add complexity to the mass models
to determine whether the allowed \hn\
range is expanded, rather than to find a better fit.

\subsubsection{The Simplest Good Fit:  C3S+DM1 (4 DOF)}
\label{c3sdm1}
A good fit to the observational constraints is possible
when
the elliptical single-power-law G1 is combined with a cluster model
containing third derivative terms.  In fact there are three branches
of solutions with acceptable $\chi^2$ values; the three solutions are
very similar save for the orientation of the third-derivative terms.
These minima are: $\chi^2/\nu$ of  $6.91/4$ ($Q=0.14$) at
$\theta_\sigma=46\arcdeg$; $6.05/4$ ($Q=0.20$) at
$\theta_\sigma=169\arcdeg$; and $4.22/4$ ($Q=0.38$) at
$\theta_\sigma=293\arcdeg$. These three solutions appear at
120\arcdeg\ intervals in $\theta_\sigma=\theta_\delta$, suggesting
that it is the addition of the $\delta$ term in
Equation~(\ref{clust2}), with its $m=3$ symmetry, that is the key in
improving the model fit so much over all previous attempts.  We will
investigate this further in \S\ref{c3dm1} below.

Without any definitive knowledge of the location of the cluster center
relative to G1, we should accept any of the three solution branches.
As mentioned in \S\ref{ccenter}, Paper I gives two weak, but
independent, lines of evidence that the cluster center lies to the
northeast of G1.  Both the {\it weak lensing mass map} and the {\it
peak galaxy density} are NE of G1 at marginal significance.  In
addition, all of the best-fit strong-lensing models orient the {\it
principal axis of the shear} (C2 term) in the range
$130\arcdeg<\theta_\gamma<150\arcdeg$.  The amplitude
$\gamma$ of the shear is 0.1--0.2, much too large to be due to
large-scale structure along the line of sight, and a strong sign that
the cluster is in fact {\it not} centered on G1.  This implies that
the cluster is centered either toward the NE or the SW of G1.  Each of
these measurements is, alone, weak evidence for a cluster center to
the NE, but they are all completely independent, so taken together
they make a more persuasive argument.  Since the 
cluster gradient direction $\theta_\sigma$ should
point roughly toward the cluster center, even for elliptical or
slightly irregular cluster masses, we should prefer the
$\theta_\sigma=169\arcdeg$ solution over the other two.  In what
follows we will examine the \hn\ allowed for
$90\arcdeg\le\theta_\sigma\le180\arcdeg$.  If $\theta_\sigma$ is left
free, the allowed \hn\ range is substantially wider.

Before we calculate constraints on $\Delta\hat t$, let us examine the
nature of this successful model for the lens mass.  Since this is the
simplest mass model which yields a good fit to the observations, we
will refer to it as the ``Best Fit Model.''
The parameters of
the successful models are listed in Table~\ref{tc3s}.  The G1 mass is again
somewhat shallower than isothermal at $\alpha=-0.83$ and, at $e=0.42$, is
about as flattened as the outermost isophotes of G1 (Paper II).  
The $PA=168\arcdeg$ differs significantly from the orientation of the
isophotes, which twist from 135\arcdeg\ to 150\arcdeg\ (Paper II).
The agreement between mass model shape and isophote shape is not
perfect, but is plausible.  The $\approx20\arcdeg$ misalignment
between mass and isophotes is only slightly larger than the typical
10\arcdeg\ misalignment found in the models of 17 lenses by Keeton,
Kochanek, \& Falco (1998).

\begin{table}
\label{tc3s}
\epsscale{1.0}
\plotone{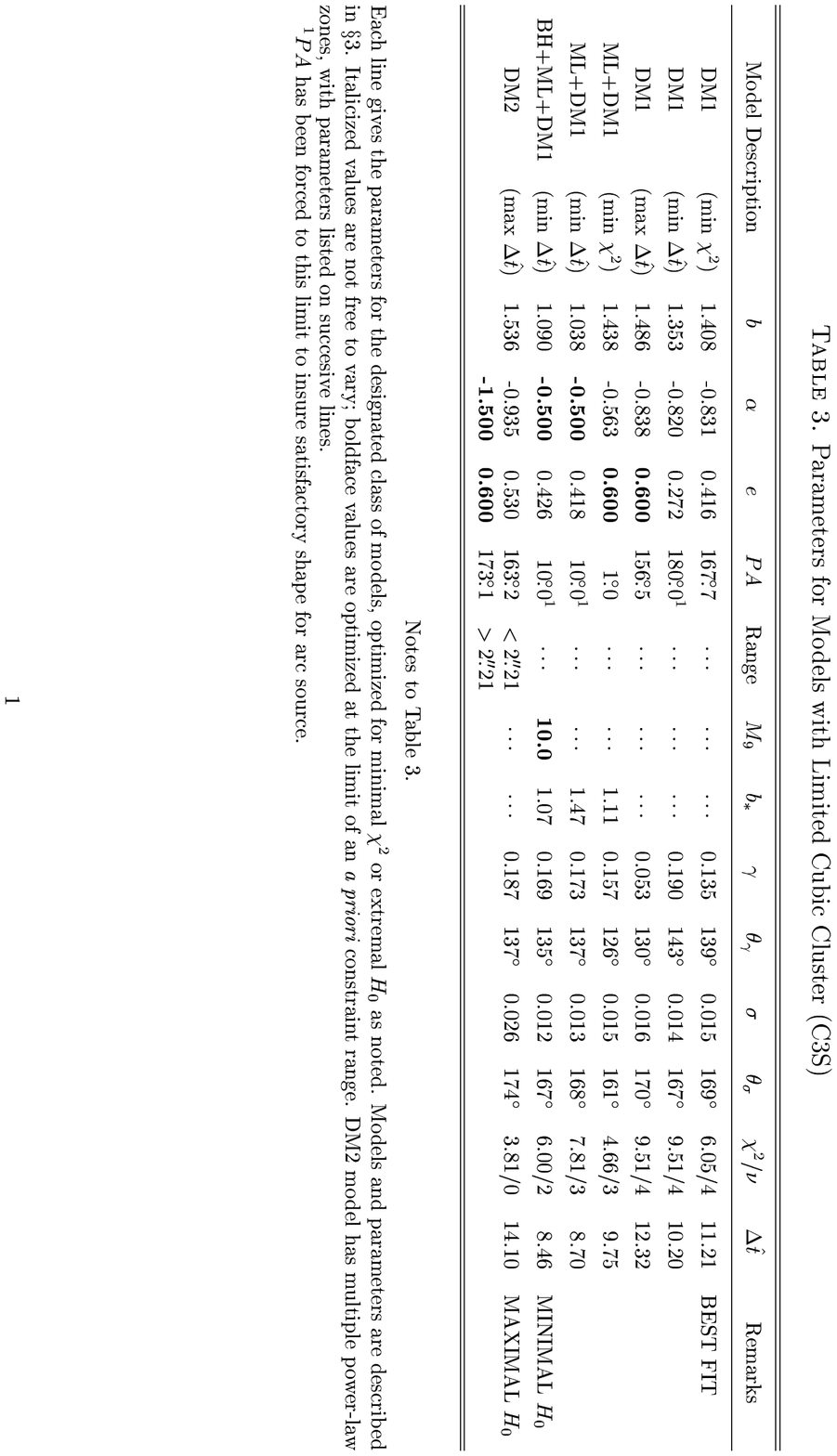}
\end{table}

The external shear axis is $\theta_\gamma=135\arcdeg$ while the third
derivative term has axis $\theta_\sigma=\theta_\delta=162\arcdeg$.  The
30\arcdeg\ misalignment is well within the range attributable to an
elliptical or irregular cluster shape.  The amplitudes of the
shear and third derivative are all reasonable given the measured
$\kappa\approx0.3$ of the cluster at G1 (see \S\ref{weak}).

\begin{figure}[t]
\plotone{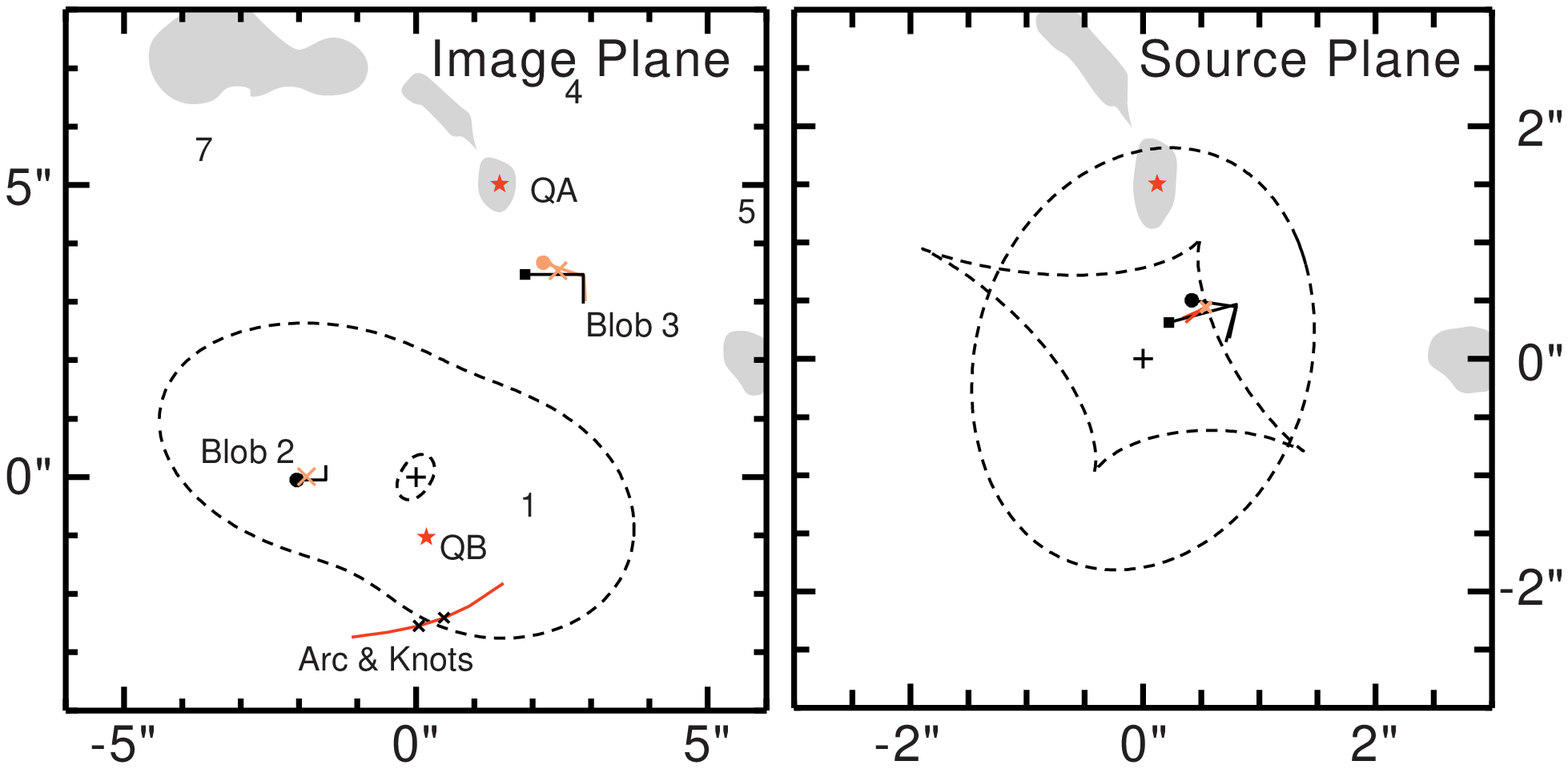}
\caption[dummy]{\small
Image (left) and source (right) planes of the 0957+561 lens system
under the best-fit C3S+DM1 model described in \S\ref{c3sdm1}.  Stars
mark the positions of the quasar images and source, with the shaded
regions outlining the VLA jets.  The crosshairs mark the position of
the G1 center, and the dashed lines denote the critical lines and
caustics of the lens.  The source of Blob 2 is schematically
illustrated as an ``L'' shape with a circle at one terminus, whereas Blob 3
is an ``L'' terminated by a square.  The corner of each ``L'' locates
the centroid of its Blob.  The arc is a light line segment,
including the two Knots ($\times$ symbols).  A zoomed view of the Blob
and arc sources is given in the middle panel of
Figure~\ref{blobsrc}. In the Image Plane, the
predicted locations for two counterimages of the Knots are marked as
light crosses atop Blobs 2 and 3; the predicted counterimage of Blob 2
is shown also atop Blob 3.  The sources for both Blobs and Knots are
astride the caustic, with the arc source being very compact.  A highly
demagnified quasar image is predicted near G1 but not shown here.
Numerals in the image plane denote the other HST Blobs, which are not
multiply imaged.
}
\label{caustics}
\end{figure}

The image and source plane geometries for this model are shown in
Figure~\ref{caustics}. 
As discussed earlier, the source for the arc is quite close
to the source for Blobs 2 and 3, and the arc is part of a 4-image
system with additional images adjacent to Blobs 2 and 3.  The source
for the arc is compact and ``folded'' back upon itself as we expected.
The source
of the Knots lies just inside the inner caustic of the lens.  The
quasar source lies inside the outer caustic; the VLA jet sources
are outside the caustics and are only singly imaged.  The third image
of the quasar appears 140 times fainter than B, even without a
``capture'' by a star.
This model is an excellent fit to the observations and not
astrophysically peculiar in any way.

With a good model in hand we now can proceed to bound the acceptable
range in $\Delta\hat t$.  With $\theta_\sigma$ restricted to the NE
quadrant, the best-fit model has $\Delta\hat t=10.87$ and the widest
range within the 95\% CL bounds ($\chi^2<9.48$ for 4 DOF) is 
$9.89<\Delta\hat t< 11.94$. 
The parameters which
extremize $\Delta\hat t$ are listed in Table~\ref{tc3s}.  

The highest value for $\Delta\hat t$ requires a G1 ellipticity set to the
{\it a priori} upper bound of $e=0.6$ (axis ratio 2:1).  Removing this
restriction results in an upper bound on $\Delta\hat t$ only 0.3\%
higher (at $e=0.62$), so the upper limit on \hn\ is not really set by
the 2:1 axis ratio criterion.

\begin{figure}
\epsscale{0.3}
\plotone{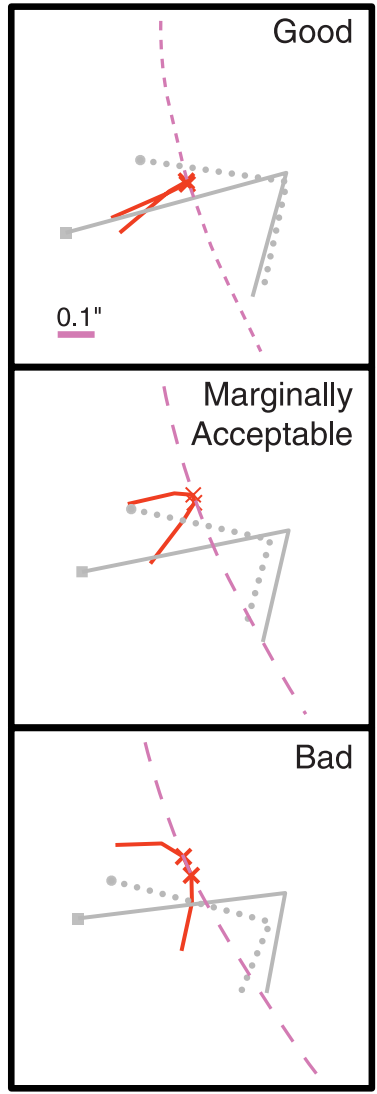}
\caption[dummy]{\small
Examples of good, marginally acceptable, and poor sources for the
arc.  The
solid grey ``L'' with square terminus is the Blob 3 source, while the
dotted ``L'' with circular terminus is the Blob 2 source.  The arc is
the ``U'' shape, and the crosses show the source positions for two
knots, generally atop the caustic (dashed line).  
We reject models for which the arc source
is more extended than in the middle panel; the bottom panel shows a
reject, while the top panel shows precisely the source behavior we
would expect.
}
\label{arcs}
\end{figure}

The search for minimal  $\Delta\hat t$ yields a
model with G1 orientation of $PA\approx40\arcdeg$, which causes the
arc source to be ``unfolded.''  Enforcing $90\arcdeg\le
PA\le180\arcdeg$ (G1 PA in NE quadrant) gives a marginally acceptable
arc source, as illustrated in Figure~\ref{arcs}.  
This is admittedly a
qualitative judgment.  In most of what follows, the models which
define the lower bound on $\Delta\hat t$ have the same problem of a
poor arc source, and we implement the G1 $PA$ restriction to enforce a
compact arc source.

\subsubsection{Adding Cluster Freedom: C3+DM1 (2--4 DOF)}
\label{c3dm1}
We explore the sensitivity of the allowable $\Delta\hat t$ 
range to the details of the cluster third derivatives by changing
the restrictions
in Equation~(\ref{c3sis}) that relate the third-derivative cluster
terms.  We discover that {\it our results are insensitive to the relations
assumed among the third-derivative terms.}  This means that our
results should be robust to ellipticity or irregularity in the
shape of the cluster dark matter.

We first set $\sigma=0$.  The best-fit model has
$\chi^2/\nu=5.57/4$ ($Q=0.23$), with a 95\% CL range ($\chi^2\le9.48$)
of $9.88<\Delta\hat t<11.97$.  This is nearly identical to the
$\Delta\hat t$ range for the previous model (C3S+DM1).

Next we set $\delta=0$.  The best-fit model has $\chi^2/\nu=7.27/4$ 
($Q=0.12$), with a 95\% CL range ($\chi^2\le9.48$)
of $9.84<\Delta\hat t<12.37$ (assuming $\theta_\sigma$ in the NE quadrant).

Adding the $m=3$ term ($\delta$) yields a somewhat better fit than
does the $m=1$ ($\sigma$) term.  But
adding {\it either} term to the lens model vastly improves the fit over the
quadratic-order cluster models.  Without these C3 terms, all
of the lensing potential has even-$m$ symmetry about G1.  It seems
that the key to a satisfactory fit is an odd-$m$ term to break the
inversion symmetry in the potential. 

We can give the cluster more freedom by allowing all four parameters
of the third-order cluster potential to be free, leaving 2 DOF for the
fit.  A marginally acceptable fit is attainable
($\chi^2/\nu=5.57/2$), and models within the 95\% CL level
($\chi^2\le6.00$) produce $10.44\le\Delta\hat t\le11.31$.

To summarize, varying the restrictions on the cluster third
derivatives does not substantially alter the range of \hn\ compatible
with the observations.  Among all C3S+DM1 and C3+DM1
models, the time delay is limited to $9.84<\Delta\hat t<12.34$,
assuming that the cluster gradient runs to the NE quadrant.

\subsubsection{Adding a Stellar Contribution: C3S+ML+DM1 (3 DOF)}
Stellar mass is certainly present in G1 so we would like to find good
lens models incorporating the ML terms.  Indeed a fit with
$\chi^2/\nu=4.66/3$ ($Q=0.20$) is found, with $b_\ast=1.11$.  It is
comforting that the best-fit value of the mass-traces-light component
is quite close to that expected from the stellar population.
Furthermore, inclusion of the stellar mass component has not degraded
the quality of the fit from the simpler ``Best Fit'' model of
\S\ref{c3sdm1}---but it {\it does} yield a substantially lower time
delay at $\Delta\hat t=9.75$ (Table~\ref{tc3s}).
The DM1 ``halo'' component is very shallow ($\alpha=-0.56$) and elliptical
($e=0.6$ at $PA=180\arcdeg$) when the ML term is included.

Because the ML term has a strong central mass peak, the dark matter
power-law index becomes shallow and the derived $\Delta\hat t$ is
lowered.  The lower limit to $\Delta\hat t$ is substantially decreased
by including the ML terms:  $\Delta\hat t=8.70$ is possible at the
95\% CL limit of $\chi^2=7.81$.  This extremal model has the halo
index at our {\it a priori} limit of $\alpha=-0.5$.  Shallower values
of $\alpha$ would produce even lower \hn\ values with good fits to the
lensing geometry.

Addition of a finite core radius to the dark matter (C3S+ML+CORE)
results in fits optimized at zero core radius, and no expansion of the
allowed range of $\Delta\hat t$.

\subsubsection{Adding Central Black Hole:  C3S+BH+ML+DM1 (2 DOF)}
\label{minh0}
\begin{figure}
\epsscale{0.8}
\plotone{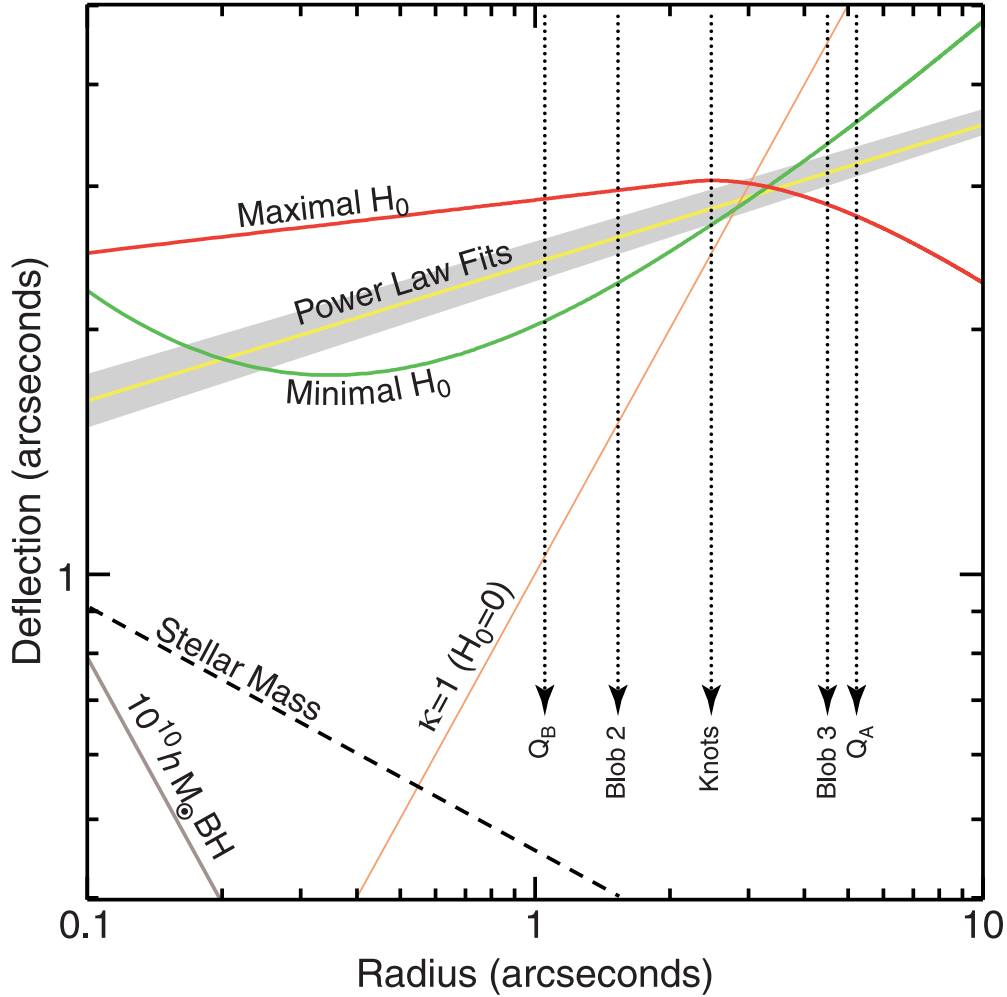}
\caption[dummy]{\small
Radial mass profiles of several models for the 0957+561 lens.  The
vertical axis is the effective deflection angle $M(<R)/\pi
R$ vs $R$, with
$M(<R)$ being the mass (in critical units)
projected within radius $r$ of the center of G1.  An isothermal
profile is {\it horizontal} in this plot. 
The 95\% CL range of power-law 
galaxy models (C3S+DM1, \S\ref{c3sdm1}) is denoted by the gray band,
and the best-fit model (with $\Delta \hat t=10.9$) runs along this
band.  A much broader range of 
\hn\ values is found if we allow the mass profile to be more complex
than a power law.  The ``Maximal \hn'' model ($\Delta \hat t=14.1$)
has two power-law segments (C3S+DM2, \S\ref{c3sdm2}), and the
``Minimal \hn'' model ($\Delta \hat t=8.5$) has power-law dark matter
with black hole and stellar mass terms added (C3S+BH+ML+DM1,
\S\ref{minh0}).  The deflection due to a $10^{10}h^{-1}M_\odot$ black hole
and to the expected stellar mass are shown near the bottom.  These
profiles are all plotted assuming $\kthirty=0$; increasing $\kthirty$ will
drive the radial profile toward the ``$\kappa=1$'' line, and also
drives the derived \hn\ toward zero.
It is apparent that \hn\ is determined largely by the slope of the
mass profile between the radii of the two quasar images (as marked).
The arc/Blob objects at the marked intermediate radii help constrain
the mass profile, but it is clear that models restricted to a
single power law G1 profile do not sample the full allowed \hn\ range.
}
\label{profiles}
\end{figure}

The addition of a black hole constrained to $M_9<10$ does not
qualitatively change the fits. It does, however, allow 
$\Delta\hat t$ values as low as 8.46 within the 95\% CL contours
($\chi^2/\nu<6.00/2$), a 3\% reduction in \hn.  For this new extremal
model, the black hole mass is at its {\it a priori} upper limit, and
the dark matter halo is again as shallow as our {\it a priori} limit
of $\alpha\le-0.5$ allows.  This is the lowest value of $\Delta\hat t$
found in any of our models, so we refer to this as the ``Minimal \hn''
model. Its parameters are in Table~\ref{tc3s} and its radial profile is plotted
in Figure~\ref{profiles}.

Thus we see that the addition of the centrally concentrated stellar
and black-hole masses has {\it weakened the lower bound on \hn\ by
15\%}, an appreciable change.  The lower bound on \hn\ also depends
strongly on our {\it a priori} limit on the dark matter radial index:
adopting the stricter limit of $\alpha<-0.6$ brings the 95\% CL lower
limit on $\Delta\hat t$ back up 7\% to 9.10.

\subsubsection{Two-Segment Power Laws:  C3S+DM2 (0 DOF)}
\label{c3sdm2}
We give G1 substantially more freedom with the DM2 model.  With the
C3S cluster model we formally have zero degrees of freedom in the fit,
but we will judge these models as if they have $\nu=1$.  Allowing
such freedom expands the range of
$\Delta\hat t$ compatible with the lensing constraints.

Pushing down the lower bound of $\Delta\hat t$ we find models with
very shallow G1 mass: $\alpha_0=-0.75$, $r_{01}=0\farcs9$, and
$\alpha_1=-0.5$.  The minimum time delay value within the 95\% CL
bound of $\chi^2\le3.81$ is $\Delta\hat t=7.36$, another 15\% decrease
from the lower limit in the previous section.  This model is not,
however, astrophysically plausible, because its surface mass density
inside $r\approx0\farcs2$ is only slightly above the 
minimum expected from the
stellar mass [{\it cf.\/} Equation~(\ref{ml})], and is quite shallow.
The dark matter density would then have to be {\it decreasing} toward
the core, which we regard as unlikely.  
We can include an ML
term with $b_\ast=1$ to enforce a density everywhere at least as large as the
expected stellar contribution, and require (as usual) that the dark
matter component increase toward the center.  The model of this type with
lowest attainable $\Delta\hat t$ within
the 95\% CL contours turns out to be virtually indistinguishable from
the lowest-$\Delta\hat t$ C3S+BH+ML+DM1 model found above.
The additional freedom for G1,
therefore, does not extend the lower bound on \hn\ if we use the
stellar mass as a floor for the G1 mass.

The upper bound on $\Delta\hat t$ is raised to 14.10 by C3S+DM2 models
which let G1 break from a nearly-isothermal slope of $\alpha_0=-0.93$ to
a steeper $\alpha_1=-1.5$ outside of $r_{01}=2\farcs2$.  This bound is
13\% higher than those derived in the previous sections, and is the
``Maximal \hn'' model in our study.  The \hn\
bound is this time strongly dependent upon our {\it a priori} limit
of $\alpha>-1.5$ on the steepness of the G1 profile.  Since the Tonry
\& Franx (1998) stellar dynamics measurements only extend to
$\pm3\arcsec$ it is not likely that their observations can rule out
this mass model.  Perhaps by comparing with studies of other
ellipticals one could exclude this mass distribution for G1, but for
now we are forced to accept this as a valid model.

\subsection{Fourth-Order Cluster Models}
An extension of the approximation for the cluster potential from
quadratic order to cubic order has greatly improved the fit to the
observations.  We have seen that the $\Delta\hat t$ values are not
very sensitive to the precise form of the cubic terms, but we would
also like to see if our results on \hn\ are robust against cluster
potential terms beyond cubic order.  

\subsubsection{Peaked  Cluster: C4P+DM1 (3 DOF)}
We might expect a gentle peak in the cluster mass near G1 to decrease
the lower bound on $\Delta \hat t$ by adding a nearly-constant mass
sheet to the vicinity of the quasar images.  Including the C4P mass
term from Equation~(\ref{cr2}), however, does not lead to any
expansion of the allowed $\Delta \hat t$ range, and the best-fit
models place $\kappa_p=0$.  Significant values ($\kappa_p\gtrsim0.1$)
are strongly excluded by the strong lensing constraints.  We conclude
that the cluster mass does not have an important maximum within
$R=30\arcsec$, in mild disagreement with the weak-lensing analysis of
Paper~I.

\subsubsection{Quartic Cluster: C4S+DM1 (2 DOF)}
The C4S restricted fourth-order cluster approximation has 6 free
parameters (\S\ref{c4s}), so we will combine it with the simplest mass
model for G1, the elliptical power-law DM1.  More complex models for
G1 would leave insufficient degrees of freedom.  We limit the
amplitude $c_4$ of the quartic cluster terms to the value they would
acquire from a singular isothermal cluster located only 10\arcsec\
away from the center of G1.  Larger values than this would make the
cluster mass supercritical within the central 10\arcsec\ region, which
is clearly not the case.
We also require the quartic terms to
point roughly toward the NE, $90\arcdeg<\theta_4<200\arcdeg$. 

The C4S+DM1 model does have an acceptable best fit, with
$\chi^2/\nu=4.33/2$ ($Q=0.11$).  The parameters are similar to those
of the best-fit C3S+DM1 model.  The range of time delays attainable
within the 95\% CL region ($\chi^2\le6.0$) are
$9.71<\Delta\hat t<12.47$.  Recall from \S\ref{c3sdm1} that the
acceptable range for the C3S+DM1 models was
$9.89<\Delta\hat t<11.94$.  So the addition of the fourth-order
cluster terms does not improve the quality of the fit to the
observations, and permits an extension of the allowed $\Delta\hat t$
(and hence \hn) range of only +4\% or -2\%, a small fraction of the
total allowed range.  Unless the cluster mass has strong features on
scales of $\sim10\arcsec$ or less, further terms in the power-law
expansion of the cluster potential should have even smaller effects
on lens models.
Thus we can conclude that, given the current set
of lensing constraints, the cubic approximation to the cluster is
necessary and sufficient for the purposes of placing bounds on \hn.

\subsection{Overview of Strong-Lensing Models}
We summarize here the properties of the various lens models we have
tested against the observations.
\begin{itemize}
\item It is not possible to fit the observed lensing geometry with
astrophysically reasonable
models that use the quadratic (C2) approximation to the cluster
potential.  All such models, with up to 12 free parameters,
are excluded at $>97\%$ confidence.  There
is good agreement among various authors on the parameters of the
best-fitting simple models, though it is clear that the real lens is
not as simple as these models.
\item Allowing third-order terms in the expansion of the cluster (C3S)
potential permits a good fit to the data, even with a single power-law
galaxy (DM1).  The best-fit C3S+DM1 model has $\chi^2/\nu=6.03/4$, a
level which occurs by chance with 20\% probability.  For this
simplest, ``Best Fit'' model, $\Delta\hat t=10.9$.
\item The fit is insensitive to the details of our restrictions on
the cluster third-order terms; a good fit generally seems to need some
term that breaks the inversion symmetry of the lens.
\item C3+DM1 and C3S+DM1 models with a range $9.84\le\Delta\hat
t\le12.37$ are consistent with the observations at 95\% CL.
\item Inclusion of {\it fourth}-order cluster terms neither improves
the fit nor significantly widens the allowable $\Delta\hat t$ range.
\item Equally good fits to the data are also available when we include a mass
term that traces the light density.  The best-fitting models have
mass-to-light normalization
$b_\ast=1.1$, where $b_\ast=1$ is what we expect from the stellar
population.  This agreement is reassuring.
\item While inclusion of the stellar mass is not required to fit the
lensing geometry, it does yield significantly lower values of
$\Delta\hat t$ within the 95\% CL.  The same is true, to a lesser
extent, of a central black hole of mass $10^{10}h^{-1} M_\odot$.  The
``Minimal \hn'' C3S+BH+ML+DM1 model pushes 
the lower bound on $\Delta\hat t$ to 8.46.
\item Allowing the G1 power law to break to a steeper value
$\alpha=-1.5$ between the quasars (``Maximal \hn'' C3S+DM2 model) 
raises the {\it upper} bound on $\Delta\hat t$ to 14.10.
\end{itemize}

Thus the best-fit simple model gives $\Delta\hat t=10.9$, with a
95\% CL range of $8.46<\Delta\hat t<14.10$.  Combining with
Equation~(\ref{dt4}), we have
\begin{equation}
\label{h0a}
H_0= 104^{+31}_{-23}\,
(1-\kthirty)
\; {\rm km}\;{\rm s}^{-1}{\rm Mpc}^{-1}\; \mbox{(95\% CL)}.
\end{equation}
We end up, therefore, with a $\pm20\%$ uncertainty in \hn\ from the
lens modeling alone.  There is additional uncertainty in
$1-\kthirty$, as discussed in the next section.  Had we studied only
the simplest model that fit the data (C3S+DM1), we would have derived
a 95\% CL range of smaller than $\pm10\%$.  The additional complexity
of the BH, ML, and DM2 terms did not significantly improve the quality
of the best fit, but it does more than double the allowed range of
\hn.  Since these additional terms are astrophysically reasonable, we
have no alternative but to consider the more complex models as
yielding valid estimates of \hn.

To what extent does the result in Equation~(\ref{h0a}) depend upon our
{\it a priori} limitations on the lens mass distribution?  The
best-fit model, fortunately, does not place any of the model
parameters at the {\it a priori} bounds.  The extremal models,
however, place the radial power-law index $\alpha$ of the G1 dark
matter distribution at its {\it a priori} limits.  The lowest-\hn\
model, of type C3S+BH+ML+DM1, has $\alpha=-0.5$, and the highest-\hn\
model (C3S+DM2) has $\alpha=-1.5$ outside of $r=2\farcs2$.  The bounds
on \hn\ thus are strongly dependent on our assumptions about a
``reasonable'' galaxy profile might be.  Assuming $\alpha<-0.6$, for
example, raises our lower bound on \hn\ by 7\%, removing one third of
the error bar.

The lowest-\hn\ model also has the black hole mass at its {\it a
priori} upper limit of $10^{10}h^{-1}\,M_\odot$.  A larger value is
unlikely, however, and only weakly affects the \hn\ bound.

We reiterate as well that we have assumed that the cluster mass
gradient points toward the northeast quadrant from G1.  Relaxing this
constraint would significantly widen the allowed \hn\ range.

Figure~\ref{profiles} shows how the radial mass profile is the most
important determinant of \hn.  We plot the deflection angle $M(<R)/\pi
R$ vs $R$, where 
$M(<R)$ is the mass enclosed within a circle of radius $R$ centered on
G1.  An isothermal profile is flat in this representation, while a
shallower radial mass profile yields an upward slope.  We see that the
shallower the mass profile between the two quasars, 
the lower the \hn\ value implied.  All the
valid models cross at a point intermediate to $Q_A$ and $Q_B$,
since the sum of the deflections angles at $Q_A$ and
$Q_B$ must of course equal the 6\arcsec\ separation between them, as
discussed in \S\ref{circsym}.

\subsubsection{Relation to Previous Models}
The inclusion of cluster terms beyond quadratic order dramatically
improved the agreement between our lens models and the observed
geometry.  The models of GN, FGS, and 4 of the 5 BLFGKS models have
only a quadratic cluster, which explains why we have found a lower
$\chi^2$ value for essentially the same constraints.  The fifth model
of BLFGKS, ``FGSE+CL,'' includes an elliptical softened-core
isothermal G1 {\it and} a singular isothermal sphere galaxy cluster.
The velocity dispersion and position of the SIS cluster are free
parameters, and the cluster potential is exact, not a quadratic
approximation.  Yet the best-fit $\chi^2$ is 41 for 7 DOF, highly
excluded.

Our C3S+DM1 model fits the data much better---why?  At first glance it
might be because we have relaxed the constraint on the position of Jet
5, but in fact we obtain an equally good fit using the error bar on
the Jet 5 position given by BLFGKS.  The main reason for the improved
fit is that we have {\it not} required that the cluster have an SIS
profile.  Indeed the best-fitting model has $\theta_\gamma=139\arcdeg$
but $\theta_\sigma=169\arcdeg$, meaning that the second and third
derivatives of the cluster potential are not aligned, and hence the
cluster is not spherical.  Furthermore, we have $\gamma=0.14$ and
$\sigma=0.015$, whereas an SIS cluster would have $\sigma=\gamma^2$.
If we take our best-fit C3S+DM1 model and require alignment of
$\theta_\gamma$ with $\theta_\sigma$ to within 5\arcdeg, the $\chi^2$
value rises from 6.0 to 26.7, strongly excluding the alignment that a
spherical cluster would generate.  An singular isothermal {\it elliptical}
cluster distribution would require $e\approx0.5$ to
generate the observed 30\arcdeg\ misalignment between the mass
gradient and the shear.  We reiterate that 
our approach of continuing the power-law
approximation to the cluster potential means that we do not have to
know or fit the global shape of the cluster, just its multipole
moments about G1.  We note that it is possible to {\it measure} the
multipole moments about G1 directly from the shear pattern in a weak
lensing map (Schneider \& Bartelmann\markcite{Sc3}, 1997).  

The apparent departure of the cluster from SIS form suggests that it
is not safe to infer the $1-\kappa$ factor by assuming that
$\kappa=\gamma$, as is done in producing some of the \hn\ estimates in
BLFGKS.

\subsubsection{The Dark Matter Distribution in G1}
The strong-lensing models for G1 give detailed information on the mass
distribution in the giant elliptical galaxy G1.  All of the acceptable
lens models share a few characteristics:
\begin{itemize}
\item The dark matter distribution is shallower than isothermal within
a radius of $\approx 2\farcs5$.  The C3S+DM1 models place the radial
index of the {\it total} G1 mass at $\alpha=-0.83$.  The stellar light
distribution is significantly steeper at $-1.9\le\alpha\le-1.3$, so
when we include a mass-traces-light term in the G1 mass, we find that
the {\it remaining} mass (``dark'' matter) is forced to significantly
shallower values of $-0.6\lesssim\alpha\le-0.5$.  The highest-\hn\
models have profiles that are nearly isothermal within
$2\farcs5\approx10h^{-1}$~kpc and significantly steeper beyond this
point.
\item The total mass surface density is well above the stellar
contribution at all projected radii above 0\farcs1 (Figure~\ref{profiles}).
\item The matter distribution is highly elliptical, $0.4\lesssim
e\lesssim0.6$ (axis ratio of 1.5:1--2.0:1), more flattened than the
inner isophotes of G1.  The dark matter halo is oriented at
$160\arcdeg<PA<190\arcdeg$ (measured from East) whereas the
isophotes twist from 135\arcdeg\ to 145\arcdeg.  Thus the dark matter
is out of alignment by 15\arcdeg--45\arcdeg\ from the outer isophotes
of the galaxy.  Both the ellipticity and position angle of the light
tend toward the dark matter values at the outer isophotes (Paper II).
\end{itemize}

\section{Determination of $1-\kappa$}
\subsection{Choices of Method}
\label{weak}
With the strong-lensing constraints satisfied, we turn to resolving
the sheet-mass degeneracy in these models.  Four methods have been
used to determine the local density $\kappa_c$ of the
galaxy cluster near the G1 center:
\begin{enumerate}
\item Weak-lensing measurements can measure
the mean mass density $\kthirty$ within the region $R<30\arcsec$ of
the G1 center (Paper I).  This method has the virtues of being
non-parametric, and of measuring precisely the quantity that is needed
(the {\it projected} mass density).  The main drawback at the moment
is the undesirably large statistical uncertainty in $\kthirty$, which
we explain below.
\item The line-of-sight velocity dispersion $\sigma$ (LOSVD) of G1 can be
measured, providing an absolute normalization of the G1 mass model and
hence breaking the sheet-mass degeneracy.  The virtue of this method
is that the measurement is now quite precise, currently yielding a
95\% CL uncertainty of $\pm12\%$ on the quantity $\sigma^2$ that
scales \hn\ (Tonry \& Franx 1998).  The drawbacks of this method are
that, first, the LOSVD measures a degenerate combination of the G1
mass {\it and} the velocity anisotropy of the stars, hence the mass
uncertainty is significantly larger than the $\sigma^2$ measurement
error.  Romanowsky \& Kochanek (1998) show that incorporation of
constraints on the higher-order moments of the velocity distribution can
reduce the uncertainty on the G1 mass to $\pm16\%$ (95\% CL) {\it if
the three-dimensional structure of G1 potential is known}.  The second
drawback of the LOSVD normalization of $1-\kappa$ is that the LOSVD
is sensitive to the {\it three-dimensional} structure of the G1
potential whereas the lensing models constrain and require the {\it
two-dimensional} (projected) mass density of G1.  Romanowsky \&
Kochanek (1998) assumed a spherical G1 mass model from GN; not only
do the strong-lensing mass models now require an elliptical projected
mass, but the structure along the line of sight is unknown.
Allowing these additional freedoms in the orbit modeling
will not only greatly complicate
the interpretation of the LOSVD data, but would undoubtedly broaden
the allowed range of $1-\kappa$.  The complications of this
interpretation lead us to prefer the weak-lensing method.  The surface
density implied under the assumption of isotropic orbits does,
however, agree well with the weak lensing measurements (Paper I).
\label{losvd}
\item The velocity dispersion of the galaxies in the G1 cluster can be
used to measure the cluster mass density (Garrett \etal\markcite{Ga3}
1992; Angonin-Willaime \etal\markcite{An1} 1994).  This method
unfortunately suffers the difficulties of both methods (1) and (2)
above:  the few known members (21) of the G1 cluster poorly determine
the cluster $\sigma^2$; and the conversion of $\sigma^2$ to $1-\kappa$
at the G1 location requires many assumptions about both the
three-dimensional shape of the cluster potential and the anisotropies
of the galaxy orbits within this potential.  It is therefore unlikely
that this method will ever approach the precision of the previous two.
\item The properties of the X-ray emitting gas in the cluster
potential can be used to normalize a model for its mass (Chartas
\etal\markcite{Ch1} 1998).  This too suffers both from very poor S/N on the
relevant measured quantities (X-ray flux, temperature, and core
radius) and in the extensive parameterization and implicit assumptions
about the physical conditions of the cluster potential and gas within
it ({\it e.g.} that the cluster is a spherical $\beta$-model with gas
in hydrostatic equilibrium).  The x-ray data, even with observational
improvements, are therefore unlikely ever to provide the best
quantitative measure of the sheet mass.
\end{enumerate}

\subsection{Weak-Lensing Measurement of $\kthirty$}
In Paper I we use the ``aperture massometry'' formula of Fahlman
\etal\markcite{Fa3} (1994) to measure the mean mass density in annuli
centered on G1:
\begin{equation}
\label{massometry}
\bar\kappa_{r<r_i} - \bar\kappa_{r_i<r<r_o}
 = r_o^2 \left\langle \gamma_T / r^2 \right\rangle_{r_i<r<r_o}.
\end{equation}
This formula gives the average mass density (in critical units) inside
the inner radius $r_i$ in term of the average tangential shear
component of galaxies $\gamma_T$ in some annulus $r_i<r<r_o$.  The
formula is exact in the weak lensing limit ($\gamma_T\ll1$), and the
second term on the left-hand side reminds us that we measure
$\bar\kappa_{r<r_i}$ only relative to the mean mass density
$\bar\kappa_{r_i<r<r_o}$ of the measurement annulus.  

Ideally the data
can be used directly in Equation~(\ref{massometry}) to yield the
desired $\kthirty$ at $r_i=30\arcsec$.  In practice there are four
complications:  first, the background galaxies are not fully resolved,
and the observed shapes are driven toward the shape of the seeing
disk, squelching the lensing signal.  Second, the source galaxies are
at a variety of redshifts, hence the mean critical density (and $\kappa$)
may differ by a scaling factor from the value appropriate to the
strong-lensing sources at $z=1.41$.  In Paper I we measure these
scaling factors by creating Monte-Carlo simulations of source galaxies
with redshift, shape, and size distributions that match what is known
about the true galaxy population.  Paper I uses a model of McLeod \&
Rieke (1995) to estimate a source redshift distribution to $V=26.5$
in an $\Omega=0.1$ Universe; the seeing-corrected $\kappa$ for this
population must be increased by $\approx40\%$ to give the $\kappa$
appropriate to the $z=1.41$ strongly-lensed sources.  This scaling
factor is uncertain, we estimate, by $\approx10\%$, due to our
ignorance of the galaxy distribution and cosmic geometry.  Since the
statistical uncertainty in $\kappa$ is much larger ($\approx30\%$), we
will ignore the calibration uncertainties at this juncture.

The third complication in the use of Equation~(\ref{massometry}) is
that the lensing is not entirely in the weak limit.  Paper I
introduces an iterative approach in which the weak-lensing formula is
used to make an initial radial profile, which is
then used to apply slight corrections to Equation~(\ref{massometry})
to account for stronger lensing near the cluster center.  As can be
seen from the top panel of Figure~6 in Paper I, this introduces only a
few percent correction to $\bar\kappa_{r<30''}$, and the details
are not critical here.

The fourth complication is that the data in Paper I extend only to
$r_o=168\arcsec$, so we must estimate the mean mass density
$\bar\kappa_{30''<r<168''}$ before we get a final
$\kthirty$.  Paper I derives this correction by assuming that the
cluster follows an isothermal profile at radii beyond the outer bound
of the data.  The second (annular) term on the left-hand side,
$\bar\kappa_{30''<r<168''}$, is 22\% of the desired
$\kthirty$ term under this assumption.  In essence we are still
subject to a sheet-mass degeneracy, although this time the sheet must
be constant across a 168\arcsec\ radius instead of the much smaller
strong-lensing region.  When smoothed over this much larger circle,
the expected $\bar\kappa$ from the cluster is far smaller, and thus so
is its uncertainty.  We will ignore the uncertainty in this 22\%
correction, and take the value derived from the isothermal
approximation in Paper I.

The mean mass density measured from weak lensing can be read from
Figure~6 in Paper I (where the lower panel incorporates the small
correction for departure from weak lensing).  Correcting to the
critical density appropriate for our $z=1.41$ strong-lensing sources,
we obtain
\begin{equation}
\label{kweak}
\kthirty = 0.26\pm0.08.
\end{equation}
This is the 1-sigma random error due to the ``shape noise'' of the
intrinsic ellipticities in the finite number of background galaxies.
Since the S/N ratio on this quantity is only 3.2, we can ignore for
the time being the calibration errors relating to the source redshift
distribution and the mass density beyond $r_o=168\arcsec$.

The weak lensing data thus yield $1-\kthirty=0.74\pm0.16$ (95\% CL).
Note that the fractional error on $1-\kappa$ is smaller than the
fractional error on $\kappa$.  This helps the error
budget a bit, but the statistical error induced on \hn\ is still
$\pm22\%$ at the 95\% CL.  Combining this weak-lensing result with the
results of strong-lensing models of Equation~(\ref{h0a}) gives the
value of the Hubble constant:
\begin{equation}
\label{h0b}
H_0= 77^{+29}_{-24}
\; {\rm km}\;{\rm s}^{-1}{\rm Mpc}^{-1}\; \mbox{(95\% CL)}.
\end{equation}
We caution the reader that the probability distribution for \hn\
within these bounds is not Gaussian; for example, the best-fitting models
span the range 65--80 \hunits, with all having essentially the same
likelihood.  

\section{Prospects for Improvement}
\label{improve}
The \hn\ from the 0957 time delay derived in Equation~(\ref{h0b}) is
disappointingly imprecise, with a $\pm30\%$ 95\% CL range that
encompasses nearly all other current estimates of \hn.  This
uncertainty is due in equal parts to remaining freedom in the
strong-lensing model of the G1 mass distribution and to statistical
uncertainty in the weak-lensing measurement of $1-\kappa$ (our weak
lensing aperture of $R=30\arcsec$ was chosen to make this the case).
Substantial improvement in the precision of the \hn\ estimate will
therefore require improvements in {\it both} aspects of the problem.
We believe that imminently available data will permit these improvements.

\subsection{Tighter Strong Lensing Constraints}

We could set tighter bounds on \hn\ if we could independently
constrain the mass profile of G1.  Measurements of the LOSVD
(and other moments of the velocity) can do
this.  The analysis will not be straightforward because of the reasons
mentioned in \S\ref{losvd}.

\begin{figure}
\epsscale{0.3}
\plotone{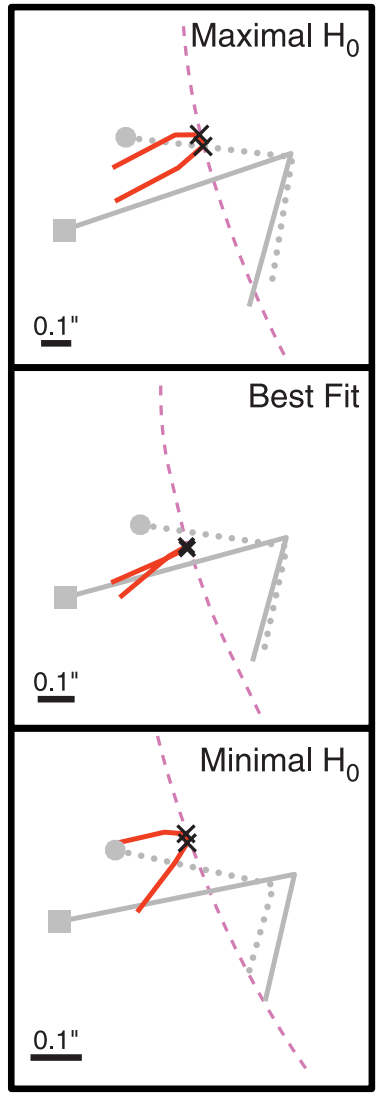}
\caption[dummy]{\small
Closeup of relative source-plane positions of Blob 2, Blob 3, and the
arc in the best-fit model and the two models which extremize \hn. 
All symbols have the same meaning as in Figure~\ref{arcs}.
Recent STIS observations of the system provide high-S/N images of
these objects, which should allow us to determine the proper
source-plane configuration.
}
\label{blobsrc}
\end{figure}

Deeper imaging of the 0957 lens system could also very easily narrow the
$\pm20\%$ modeling uncertainty in \hn.  The arc and Blobs 2 \& 3 are
images of a common source object.  The WFPC image in Paper II has
insufficient S/N to see the morphology of these objects.  A deeper
image would allow us to create a precise correspondence between
different components of these images, thus further constraining the
models.  Figure~\ref{blobsrc} shows the relative source-plane
positions of the arc and Blobs in the best-fit and \hn-extremizing
models.  We see that a detailed view of where the arc/Knot source lies
within the Blobs could distinguish these models.  Deep STIS
observations of this system obtained in March, 1999 should provide
the S/N level required.

We believe it to be quite
likely that additional multiply-imaged features will be discovered
either in our recent STIS observations of 0957, in the existing
NICMOS images of the system (Kochanek \etal\ 1998), or in future radio
imaging.  This would help pin down the radial mass profile and
decrease the \hn\ uncertainty due to the modeling.

\subsection{Tighter Weak Lensing Constraints}
The weak-lensing measure of $\kthirty$ is limited primarily by
statistical fluctuations in the shapes of the observed galaxies.
Improvement is possible if an image of the field can be obtained with
a higher density $n_g$ of resolved galaxies at S/N$\gtrsim10$.   The
uncertainty in $\kthirty$ scales as $n_g^{-1/2}$. The
$\kthirty$ measurement in Paper I uses 67 galaxies per square arcmin
at $24<V<26.5$, many of which are poorly resolved even in the 
excellent seeing (FWHM of 0\farcs6) of the CFHT image. 
The image is
not particularly deep, having been obtained with a front-illuminated
CCD; a higher $n_g$ over a larger field could be obtained with
currently-available thinned CCDs.  Even more promising is the use of
the Hubble Space Telescope; even relatively short (3600 s) exposures with the
WFPC2 camera yield arclet densities of $n_g\approx50\;{\rm
arcmin}^{-2}$ (Hoekstra \etal\markcite{Ho1} 1998), with essentially
all being completely resolved.  A mosaic of deeper WFPC2 images, or, even
better, a few images from the upcoming Advanced Camera imager, would
allow a large increase in $n_g$ and hence in the precision
of $\kthirty$.

Improved S/N in the weak lensing data will also help refine the strong
lensing model, since the arclets can be used to quantify the multipole
moments of the mass distribution both internal and external to the
$R=30\arcsec$ division that we have placed between the weak and strong
lensing regions (Schneider \& Bartelmann 1997).  The present data are not
sufficiently accurate to yield useful constraints beyond the monopole
moment $\kthirty$.

\section{Conclusions}
\label{conclusion}
We have pursued the goal of an independent, global, conceptually simple
measure of \hn\ via the
time delay of the gravitational lens 0957+561.  The simplicity is lost,
however, as we delve into the messy consideration of realistic models
for the mass distribution in the lens.  Despite the accuracy of the
time delay measurement, indeterminacy in the G1 mass model and the
cluster mass each add $\pm20\%$ uncertainty to the value of \hn, with
our final determination being 
$H_0= 77^{+29}_{-24}$\hunits\ (95\% CL).  While the resultant
accuracy of our \hn\ value is disappointing, the good news is that we
have, for the first time, produced a lens model that accurately
reproduces the many observed features of this lens.  

Our investigation
of an extensive variety of lens models shows that
the G1 lens {\it must} be elliptical and {\it must} be shallower
than isothermal, and the surrounding cluster {\it must} be
considered in more detail than a constant magnification and shear, and
the cluster {\it must} depart from circular symmetry.
The G1 lens mass profile {\it can} be a simple power law, but a
much wider range of \hn\ values are accessible if we allow the mass
profile the freedom to depart from a single power law.  In particular,
the inclusion of a mass component that traces the light in G1, with a
M/L ratio as expected from the stellar population, leads to a lowered
value of \hn.  
The range of allowed \hn\ depends strongly on the {\it a priori}
limitations placed upon the G1 mass profile.  

Constraining the lens with only the two-image system (quasars and
jets) gives an extremely wide range of possible \hn\ values (Kochanek,
1991) unless one artificially restricts the G1 mass distribution to
circular or isothermal profiles.  The 0957 system now also includes
a likely four-image system (arc \& Blobs) at a variety of radii from the
G1 center, which reduces the \hn\ uncertainty from modelling to
$\pm20\%$ under our broad range of possible G1 models.
Additional multiply-imaged sources would curtail this freedom
substantially.

The mass-sheet degeneracy from the surrounding cluster
is broken in a straightforward,
non-parametric fashion with the weak lensing data of Paper I.  Improvements
to this measurement require only deeper high-resolution images.  LOSVD
measurements of G1 can also break the mass-sheet degeneracy, but this
will require dynamical modelling of realistic ({\it e.g.} elliptical,
non-isothermal) potentials.

We believe that the uncertainties on \hn\ from this system will be cut
in half within a year or so.  But 0957+561 is now only one of three
gravitational lenses useful for \hn\ determination:
PG~1115+080 (Schechter \etal\markcite{Sc1} 1997,
Barkana\markcite{Ba2} 1997) and B0218+357 (Biggs \etal\markcite{Bi1}
1999) both now have accurately determined time delays, and known
redshifts for both source and lens galaxies.  Resultant values of \hn\
have already been published for each:  $H_0=44\pm8$ (isothermal lens) or
$H_0=65\pm10$ (mass-traces-light lens) for PG1115 (Impey
\etal\markcite{Im2} 1998), and $H_0=69^{+13}_{-19}$ for B0218 (Biggs
\etal\ 1999), each at 95\% CL in the usual units.  While these lenses
would seem to offer higher precision on \hn\ than the result derived
for 0957 in this paper, {\it we believe that the claimed precision on
\hn\ is overestimated} because the lenses have as yet been modelled only
with relatively simple mass distributions---isothermal ellipsoids.
While the lens observations may be well fit by these simple models,
this does not exclude the possibility that the mass distribution may
take other sensible forms.  In the case of 0957, we found the lens
well fit by the C3S+DM1 power-law ellipsoid, but the allowed \hn\
range is much larger when we allow additions such as the
mass-traces-light component.

It has been claimed that the 4-image systems such as PG1115 will offer
better \hn\ constraints than 2-image systems such as 0957 or B0218.
We have seen, however, that the 0957 model is now constrained by a
2-image {\it and} a 4-image system and still has 20\% uncertainty in
\hn\ from the G1 mass model.  A competitive measure of \hn\ will
apparently require even more constraint on the lensing geometry.
For 0957+561, it is likely that other multiply-imaged background
sources will be found in deep images, but this is less likely for the
other systems simply because their strong-lensing regions are smaller
(Einstein-ring radii of 3\arcsec\ for 0957, 1\farcs1 for PG1115, and
0\farcs17 for B0218).  Fortunately there is hope because
{\it all three} systems now show ring-like images of the quasar host
system:  0957 and PG1115 from NICMOS images (Impey \etal\ 1998; 
Kochanek \etal\ 1998)
and B0218 in radio images (Patnaik \etal\markcite{Pa1} 1993).  In
theory, a high-S/N image of a well-resolved ring with extended
structure can offer very strong constraints on the mass model
(see method of Wallington \etal\markcite{Wa1} 1996).  
At present, however, such information has not yet
been used in detailed modelling of 
any of these systems (indeed the S/N of present
ring images may be insufficient).  So we reiterate our caution:  the
0957 system currently has {\it many} more quantitative constraints
on lensing geometry than either PG1115 or B0218, and yet a realistic
examination of possible mass distributions leaves \hn\ uncertain by
$\pm20\%$, even ignoring the uncertainties in $1-\kappa$.

Half of our \hn\ error budget arises from uncertainties in $1-\kappa$,
which is a substantial correction for 0957 because the lens is in a
modest cluster.  Will other lenses be free of errors arising from the
mass-sheet degeneracy?  In the case of PG1115, it is clear that the
lens galaxy is in a group, and the influence of the group potential is
important for the lens model (Impey \etal\ 1998), so the mass-sheet
term (and higher-order ``cluster'' potential terms) must be considered
in a thorough examination of the lens model.  We note that the
precision of weak lensing and LOSVD measurements in determining
$\kappa$ are {\it independent of the value of $\kappa$}, so the mass
sheet problem may be no more accurately resolved in PG1115 than in
0957, despite the fact that the mass sheet is weaker.  For B0218,
which appears to be an isolated spiral, there may be {\it a priori}
reasons why $\kappa$ can be assumed to be close to zero, and it need
not be measured accurately by weak lensing or LOSVD.  

We have introduced unpleasant but realistic complexity and uncertainty
into the determination of \hn\ via gravitational time delays.  We
consider this a sign of the maturity of the field rather than a
discouragement of further investigation, though, since we believe that
each lens will be left with $\approx10-15\%$ uncertainty in \hn\ once
detailed lens models incorporating a variety of lensed features are
produced for each lens.  Since the errors in each case are
independent, the increasing numbers of systems with known time delays
will continue to yield a more accurate average \hn\ value.  We
reiterate the virtues of the lensing \hn: it is a global measurement
of geometry, not affected by local anomalies of flow or expansion; it
is physically based, so that it is difficult to imagine any undetected
systematic
effects that might affect its results, aside from a failure of General
Relativity.  Finally, it is completely independent of other methods,
and the continued agreement with these other methods increases confidence
in the entire distance ladder.

\acknowledgements
We would like to thank J. A. Tyson and G. Rhee for their collaboration
during other phases of this investigation.  M. Garrett provided
detailed data and discussion on his observations of the VLBI jets.  We
further thank C. Kochanek and the CASTLeS group for promptly showing
us the NICMOS images of 0957 from their survey.  We would also like
to thank the anonymous referee for an impressively prompt and thorough
review of this lengthy paper.
Support for this work
was provided by NASA through grants \#HF-01069.01-94A (PF) and
\#GO-05979.0X-94A (GB) from the Space Telescope Science Institute,
which is operated by the Association of Universities for Research in
Astronomy Inc., under NASA contract NAS5-26555.  GB is further
supported by grant \#AST-9624592 from the National Science Foundation.

\appendix
\section{Justifications for De-Emphasizing Jet Precision}
\label{jets}
The positions of components A5 and B5 are determined to
high precision, with formal uncertainties of only $\approx0.1$~mas,
or 0.2\% of their
displacements from the quasar cores. Use of these tiny uncertainties
in our fitting procedure would effectively constrain the local magnification
matrices (second derivatives of the lens potential $\phi$) to 0.2\%.
We choose to relax this constraint by increasing the sizes of
the error ellipses for the jet image positions to circles
with radii equal to 1\% of their
displacements from the quasar cores.  There are several justifications
for this:
\begin{itemize}
\item 
The 0.1~mas formal uncertainties given by the VLBI analyses may
be underestimated. The A5/B5 positions
change by $\approx0.3$~mas for BLFGKS's various modelling procedures,
which we take as more indicative of the uncertainty.
The complex structures of the jets are being fit by a simplified model
of a chain of Gaussians.  The fitted locations of these idealized jet
components probably is somewhat dependent upon the $uv$-plane coverage
({\it e.g.\/} the effective resolution) of the observations, and
perhaps the S/N level.  Since the B jet is a demagnified version of
the A jet, it is effectively being viewed with different $uv$-plane
coverage (relative to its angular size) than the A jet.  It might
therefore be imprudent to trust the relative positions of the A and B
Gaussian components to the formal 0.1~mas error.
\item
The quantity of interest to us is the gravitational potential
difference between quasars A and B, which is a large-scale property of
the lens.  Requiring jets A5 and B5 to have a common source is in
essence a constraint upon the magnification near the A and B quasars,
which depends primarily on the lens mass density at A and B image locations. 
Specifying such a local property as tightly as
0.1\%
could misleadingly constrain the global
solution for the G1 mass parameters, if the parameterized form of the
mass distribution does not exactly match the true distribution.
It is unlikely that any simple parametric model of the G1 mass
would describe the true mass distribution to better than 1\% accuracy
on 50~mas scales.  We consider it prudent, therefore, to increase the
uncertainties in jet positions, thereby allowing the models some
freedom in 
local magnification, which allows for the possibility that the
true surface density has small-scale deviations from our model.
\item
Increasing the error ellipse on the jet components can only increase
the range of \hn\ values from models with acceptable $\chi^2$ values.
Our approach is therefore the conservative one with regard to
constraining \hn.
\end{itemize}

\section{Potentials for Elliptical Power-Law Mass Distributions}
\label{mpole}
We wish to calculate the potentials and its derivatives for power-law
elliptical mass distributions as described by Equations~(\ref{ellip1}).
We choose to expand the mass distribution into a multipole series,
each term of which has easily derived and rapidly calculable potential
and derivatives.  We define
\begin{eqnarray}
\label{expand}
\Sigma(r,\theta)/\Sigma_{\rm crit} & = & 
	br^{\alpha}\left[ 1-e\cos2(\theta-PA)\right]^{\alpha/2}
	(1-e^2)^{-\alpha/4} \nonumber \\
 & \equiv &
    \sum_{m=0}^{\infty} a_m(e,\alpha) r^\alpha (1-e^2)^{-\alpha/4}
	\cos\left[2m(\theta-PA)\right] \\
 \Rightarrow \qquad a_m(e,\alpha) & = & 
    {2b \over \pi} \int_0^\pi
	(1-e\cos2\theta)^{\alpha/2}\,\cos2m\theta\,d\theta \nonumber
	\qquad (m>0).
\end{eqnarray}
Expanding the parentheses under the integral as a power series in $e$
we obtain
\begin{equation}
\qquad a_m(e,\alpha)  = 
	{2b \over \pi} \sum_{j=0}^\infty 
	\left[ { {(-e)^j} \over {j!}}
	{ {(\alpha/2)!} \over {(\alpha/2-j)!} }
	\int_0^\pi \cos^j2\theta\,\cos2m\theta\,d\theta
	\right] 
\end{equation}
for $m>0$.  The integral vanishes unless $j-m$ is even and positive,
in which case
\begin{equation}
\int_0^\pi \cos^j2\theta\,\cos2m\theta\,d\theta
=  { {\pi j!} \over {
	2^j\left({{j-m}\over2}\right)!\left({{j+m}\over2}\right)!} }
\end{equation}
Defining $2k\equiv j-m$ and $\gamma\equiv-\alpha/2$, 
the above two equations may be combined to give
\begin{equation}
\label{am}
a_m(e,\gamma)  = { {be^m} \over {2^{m-1}}}
	\sum_{k=0}^\infty \left({{e^2}\over 4}\right)^k
	{ {(\gamma+m+2k-1)!} \over {(\gamma-1)! k! (k+m)!} } \qquad (m>0).
\end{equation}
For $m=0$ there is an additional factor of 2.
Successive summands are easily computed by recursion and the sum
converges rapidly for small $e$.  Even for $e\gtrsim0.5$
convergence is rapid beyond $k\sim m$.  Substituting the coefficients 
Equation~(\ref{am}) into the multipole sum in Equation~(\ref{expand})
gives our expression for the power-law ellipse surface density.  In
practice we find that carrying multipoles up to $m=3$ and the
expansion of multipole coefficients to order $e^6$ describes the
surface density to 1\% accuracy even at our most difficult limits of
$\alpha=-1.9$ and $e=0.6$.

With the density expanded as multipole moments, the solution for the
potential is straightforward.  We assume that the mass distribution is
described by Equation~(\ref{expand}) between radii $r_-$ and $r_+$.
Each term in the dimensionless surface density of the form
\begin{equation}
a_m r^\alpha \cos[2m(\theta-PA)]
\end{equation}
gives rise to a term in the potential of the form
\begin{equation}
{ {a_m} \over {2m} }\cos[2m(\theta-PA)] \left\lbrace
\begin{array}{rrrc}
{ {r_-^{\alpha+2-2m}-r_+^{\alpha+2-2m}} \over {\alpha+2-2m} }  r^{2m}
 & & & r<r_- \\[3pt]
{ {-r_+^{\alpha+2-2m}} \over {\alpha+2-2m} } r^{2m}
 & + \; { {4m} \over {(\alpha+2)^2-4m^2} } r^{\alpha+2} 
 & + \; { {r_-^{\alpha+2+2m}} \over {\alpha+2+2m} } r^{-2m}
 & r_-<r<r_+ \\[3pt]
 & \multicolumn{2}{r} {
 { {r_-^{\alpha+2+2m}-r_+^{\alpha+2+2m}} \over {\alpha+2+2m} } r^{-2m}
 }
 & r>r_+ 
\end{array}
\right.
\end{equation}
This equation breaks down for $\pm2m=2+\alpha$, but if we restrict
our galaxy profiles to $-2<\alpha<0$ there will be no difficulties.
The analytic derivatives of $\phi$ with respect to the $x$ and $y$ coordinates
are then easily calculated.

\section{Surface Density of Stellar Component of G1}
\label{mlnorm}
Equation~(\ref{ml}) describes a surface mass density for G1 that
traces its observed surface brightness distribution.  Here we derive
the scale factor between the mass and luminosity distributions that we
should expect for an old stellar population.

The observed F555W ($=V$) filter band has
a central wavelength of 408~nm in the $z=0.36$ rest frame of G1, very
close to the nominal $B$-band central wavelength of 445~nm.  We can
therefore estimate $S\!B_{B,0}$, the rest-frame B-band surface
brightness, and be insensitive to assumptions about the source
spectrum.  This is given by
\begin{equation}
\label{sb}
S\!B_{B,0} = S\!B_V - K_V(z=0.36) + (B-V)_0 - 10\log(1+z),
\end{equation}
where $S\!B_V$ is the observed $V$-band surface brightness, $K_V$ is the
$K$-correction in $V$ band for redshifting the spectrum to $z=0.36$
(as defined in Coleman, Wu, \& Weedman\markcite{Co1} 1980),
and $(B-V)_0$ is the color the galaxy would have if viewed in its rest
frame.  For the spectrum of a present-day elliptical galaxy,
$K_V(z=0.36) + (B-V)_0 = -0.16$ from values in Coleman, Wu, \& Weedman
(1980) or $-0.14$ from values in Fukugita \etal\markcite{Fu1}
(1995).  As expected these agree quite well since the transformation
is nearly independent of source spectrum, and we will take
$K_V(z=0.36) + (B-V)_0 = -0.15$ for the G1 rest-frame spectrum (which
will be of a younger population than a present-day elliptical).

Given the rest-frame surface brightness, the surface mass density can
be expressed in terms of the $B$-band mass-to-light ratio (in solar
units) $\Phi_B$:
\begin{equation}
\label{massd}
\Sigma_\ast = \Phi_B 10^{-0.4(S\!B_{B,0}-B_\odot)} 
	{ {M_\odot} \over { (10 \, {\rm pc})^2 }}
	\left( { {1 \,{\rm rad}} \over {1\arcsec}} \right)^2.
\end{equation}
Combining Equations~(\ref{scrit}), (\ref{sb}), and (\ref{massd}),
$B_\odot=5.51$, and the above-mentioned color and $K$ corrections
yields 
\begin{eqnarray}
\label{ml2}
{ \Sigma_\ast \over {\Sigma_{\rm crit}}} & = &
	\Phi_B 10^{-0.4(S\!B_V-  K_V(z=0.36) + (B-V)_0-B_\odot)}
	(1+z)^4
	{ {4\pi G M_\odot} \over {c^2}}
	{ {D_{OL}D_{LS}}  \over {D_{OS} (10\,{\rm AU})^2}} \nonumber\\
   &  = & \Phi_B h^{-1} 10^{-0.4(S\!B_V-19.46)} .
\end{eqnarray}

We have as usual taken $H_0 = 100h\,{\rm km\,s^{-1}\,Mpc^{-1}}$, and
have assumed $\Omega=1$, $\Lambda=0$.  For an open cosmology the
answer is 10\% higher, while it can rise by 60\% for a
$\Lambda$-dominated Universe.
 
What value is expected for $\Phi_B$?  The synthesized old stellar
populations of Worthey \markcite{Wo1}(1994) with $[{\rm Fe}/{\rm
H}]=0.25$ have $\Phi_B=0.62\,T_{\rm Gyr}$ to within a few percent
(with $T_{\rm Gyr}$ the age of the Universe in Gyr).  Changing the
metallicity to 0 or 0.5 changes $\Phi_B$ by -25\% or +10\%,
respectively, so we will place a $\pm20\%$ uncertainty on this value.
In an $\Omega=1$ cosmology, the age of the Universe at $z=0.36$ is
$4.2h^{-1}$~Gyr.  This age can be higher for open or
$\Lambda$-dominated Universes.  The resultant estimate for $\Phi_B$ of
the G1 stellar population is $\Phi_B=2.6h^{-1}\pm20\%$ ($\Omega=1$).
Using $S\!B_V=22.4$ at $r=1\arcsec$ (Paper II), we obtain
\begin{eqnarray}
{ \Sigma_\ast \over {\Sigma_{\rm crit}}} & = &
	  10^{-0.4(S\!B_V-20.5)}h^{-2} \nonumber\\
 & = & (0.17\pm0.03)h^{-2} \qquad (r=1\arcsec,\,\Omega=1).
\end{eqnarray}
The highest anticipated $h$ is 0.8 , 
and we must remember that the mass density in
Equation~(\ref{ml}) will be scaled by $1-\kappa_c\approx0.7$
[Equation~(\ref{kap}) and \S\ref{weak}].  We then arrive at the
prefactor of 0.16 in Equation~(\ref{ml}) as the value we expect from
the stellar population of G1.  This minimum value might be lowered if
the mean metallicity is solar or below; the minimum value can be up to
twice as large if the Universe is dominated by a cosmological constant
or if $h=0.5$.

{}

\end{document}